\documentclass[11pt]{article}
\usepackage{xspace}
\usepackage{graphicx}
\usepackage{amsmath}
\usepackage{amssymb}
\usepackage{color}

\definecolor{Red}{rgb}{1,0,0}
\definecolor{Green}{rgb}{0,1,0}
\definecolor{Blue}{rgb}{0,0,1}
\definecolor{Black}{rgb}{0,0,0}

\def\beq{\begin{equation}}
\def\eeq#1{\label{#1}\end{equation}}
\def\eeqn{\end{equation}}

\def\beqa{\begin{eqnarray}}
\def\eeqa#1{\label{#1}\end{eqnarray}}
\def\eeqan{\end{eqnarray}}

\let\bar=\overbar

\def\etal{{\it et al.}}

\def\Dslash{\not{\hbox{\kern-4pt $D$}}}
\def\dslash{\not{\hbox{\kern-2pt $\del$}}}

\def\msb{{\bar{\ssstyle M \kern -1pt S}}}

\usepackage[version=3]{mhchem}
\usepackage{subfigure}
\usepackage{wrapfig}
\usepackage{lineno}

\newcommand{\CP}{$CP$}
\newcommand{\ko}{K^0}
\newcommand{\kobar}{\overline{K}^0}
\newcommand{\klpipi}{K_L \to \pi^+\pi^-}
\newcommand{\klpiopio}{K_L \to \pi^0\pi^0}
\newcommand{\klpiopiopio}{K_L \to \pi^0\pi^0\pi^0}
\newcommand{\epoe}{\epsilon'/\epsilon}
\newcommand{\reepoe}{Re(\epoe)}
\newcommand{\kpinn}{K \to \pi \nu \overline{\nu}}
\newcommand{\klpionn}{K_L \to \pi^0 \nu \overline{\nu}}
\newcommand{\kppipnn}{K^+ \to \pi^+ \nu \overline{\nu}}

\newcommand{\sm}{standard model}
\newcommand{\copyrightAPS}{{\footnotesize \textcircled{c} 2014 The American Physical Society.}}

\newcommand{\physrev}[4]{#1, Phys.\ Rev.\ {\bf #2}, #3 (#4).}
\newcommand{\prl}[4]{#1, Phys.\ Rev.\ Lett.\ {\bf #2}, #3 (#4).}
\newcommand{\prd}[4]{#1, Phys.\ Rev.\ D {\bf #2}, #3 (#4).}
\newcommand{\physlett}[4]{#1, Phys.\ Lett.\ {\bf #2}, #3 (#4).}
\newcommand{\plb}[4]{#1, Phys.\ Lett.\ B {\bf #2}, #3 (#4).}
\newcommand{\nuclphys}[4]{#1, Nucl.\ Phys.\ {\bf #2}, #3 (#4).}
\newcommand{\nim}[4]{#1, Nucl.\ Inst.\ Meth.\ {bf #2}, #3 (#4).}

\def\Title#1{\begin{center} {\Large {\bf #1} } \end{center}}

\begin{document}

\Title{\CP\ Violation in Kaon Decays (II)
%[v1.4 Dec. 16, 2014 \today]
}

\bigskip\bigskip

%+\addtocontents{toc}{{\it D. Reggiano}}
%+\label{ReggianoStart}

\begin{raggedright}  

%% Authors - you should specify at least one author as follows.
Taku Yamanaka\index{Yamanaka, T.}, {\it Osaka University}\\
%Joe Bloggs\index{Bloggs, J.}, {\it Queen Mary University of London}\\
%% In case you want to have more than one author please follow the format
%% shown below, listing the individual authors AND also making sure
%% that each author is given a unique index entry.
%Joe Another\index{Another, J.}, {\it Another University}\\
%Someone Else\index{Else, S.}, {\it Another University}\\

%% In case you want to quote your collaboration please modify the text below.
%% If this is not relevant then you may comment out the following line.
%\begin{center}\emph{On the behalf of the Bloggs Collaboration.}\end{center}
\bigskip
\end{raggedright}

{\small
\begin{flushleft}
\emph{To appear in the proceedings of the 50 years of \CP\ violation conference, 10 -- 11 July, 2014, held at Queen Mary University of London, UK.}
\end{flushleft}
}

%\linenumbers
\section{Introduction}
Major progress has been made in kaon physics in the past 50 years.
The number of observed $\klpipi$ events has increased by
6 orders of magnitude, and
the observed \CP\ violation was experimentally proven 
to be caused by a complex phase in the CKM matrix.
This mechanism is now a fundamental piece of the \sm.
Recent kaon experiments are now even searching for new physics 
\emph{beyond} the \sm\ with $\kpinn$ decays.
The branching ratios of $\kpinn$ decays are 7--8 orders of magnitude smaller than 
the branching ratio of $\klpipi$,
and  \CP-violating $\klpiopio$ decay is now a major background for
$\klpionn$ experiments.

This paper reviews 
the progress of kaon experiments in the US and Japan as requested by 
the conference organizer,
and 
how the 6--7 orders of magnitude improvements 
were possible in the past 50 years.\footnote{%
Unless noted, the years are given in published years.}

%===============================
\section{Quest for $\epoe$}
Soon after the discovery of \CP\  violation~\cite{Christenson:1964fg}, 
the $\klpipi$ decay was explained to be caused by an admixture of a
\CP-even component in the $K_L$~\cite{lee_annrev_1966}:
\begin{equation}
	|K_L\rangle \sim |K_{odd}\rangle + \epsilon |K_{even}\rangle ,
\end{equation}
where this \CP-even component was decaying to a \CP-even $\pi^+ \pi^-$ state.
This admixture is introduced by a complex phase in the $K^0 - \overline{K^0}$ mixing.
%and thus called \textit{indirect} \CP\  violation.
The next question was whether the \CP-odd $K_{odd}$ can directly decay to 
a \CP-even $\pi\pi$ state.
Such process is called \textit{direct} \CP\  violation.
If the direct \CP\  violation exists,
the ratios between decay amplitudes:
\begin{eqnarray}
	\eta_\pm \equiv &A(K_L \to \pi^+ \pi^-)/A(K_S \to \pi^+\pi^-) & = \epsilon + \epsilon' \textrm{~and}\\
	\eta_{00} \equiv &A(K_L \to \pi^0 \pi^0)/A(K_S \to \pi^0\pi^0) & = \epsilon - 2\epsilon'
\end{eqnarray}
can be different due to isospin.
The existence of the direct \CP\  violation can thus be tested by checking whether
the double ratio:
\begin{eqnarray}
	R &\equiv &\frac{BR(K_L \to \pi^+ \pi^-)/BR(K_S \to \pi^+ \pi^-)}
				{BR(K_L \to \pi^0 \pi^0)/BR(K_S \to \pi^0 \pi^0)}\\
		& = & \left| \frac{\eta_\pm}{\eta_{00}} \right|^2\\
		& \simeq &1 + 6Re(\epsilon'/\epsilon)
\end{eqnarray}
deviates from 1 or not.

The superweak model~\cite{sw} explained that a very weak unknown interaction 
that changes the strangeness by $\pm$2 brings in the phase.
However, the superweak model cannot violate \CP\  in the \(K_L \to \pi\pi\) decay process
because it cannot contribute to such a $\Delta S = \pm 1$ transition.

\subsection{Advancement in Experimental Technologies}
To measure the double ratio $R$, 
high statistics is required.
This means that 
both a higher kaon flux and detectors capable to collect data with higher rates are needed.

\subsubsection{Production Target}
\begin{figure}[!ht]
    \begin{center}
        \includegraphics[width=0.66\columnwidth]{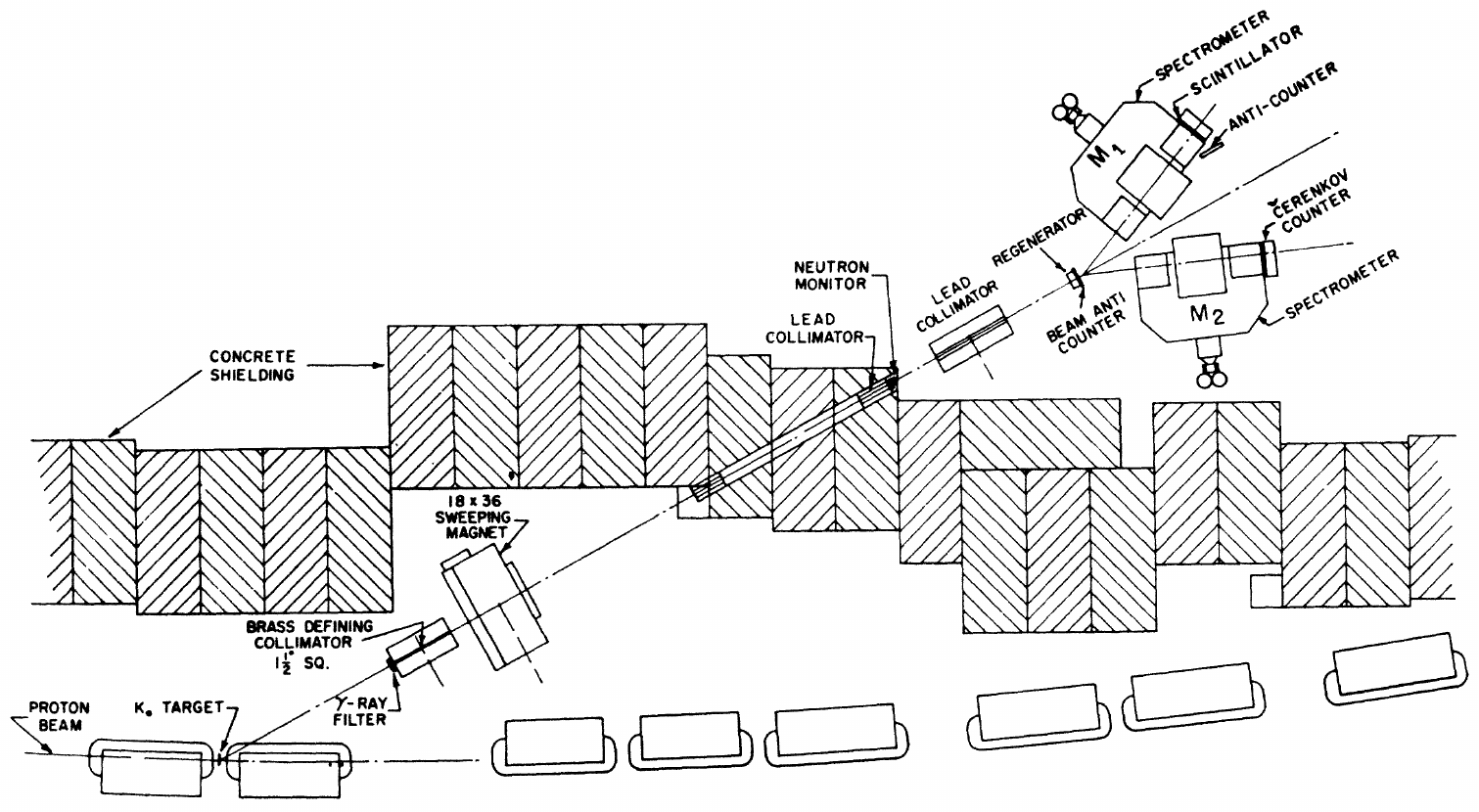}
        \caption{Plan view of the neutral kaon experiment at BNL AGS.
		For the experiment that discovered \CP\  violation,
		the regenerator was removed, and a He bag was installed between the collimator and spectrometers
		~\cite{christensen_pr140_1965}.  \copyrightAPS
		}
        \label{fig:christensen_prb140_1965_Fig1}
    \end{center}
\end{figure}

To get a higher kaon flux, the advancement of accelerators was essential,
but there was also a change in production targets.
In 1964, the experiment that first discovered \CP\  violation with 
35 $\klpipi$ events used 
an ``internal target'' in the accelerator ring as shown in Fig.~\ref{fig:christensen_prb140_1965_Fig1}.
The target was a Be wire with 0.5 mm in diameter~\cite{christensen_pr140_1965}.

In 1969, at the CERN PS, the proton beam was extracted from the accelerator to bombard
a 72-mm-thick tungsten target.
About 400 $\klpipi$ events were collected~\cite{bohm_npb9_1969},
ten times more than in the first experiment.

%-------------------------------------------------------
\subsubsection{Charged Spectrometers}
There was also a change in detector technologies.
The standard tracking devices in the 1960s were spark chambers.
They had an advantage that only the tracks of interest can be made visible
by applying HV pulses for triggered events.
Mirrors were arranged to capture sparks in multiple spark chambers
of each event in a single photograph.
For the first \CP\  violation experiment, 
tracks in the photographs were scanned by human ``scanners'' with digitized angular 
encoders~\cite{christensen_pr140_1965}.
%utilizing drawing boards and precision gears.%\cite{jim}.

\begin{wrapfigure}{r}{0.45\columnwidth}
	 \includegraphics[width=0.45\columnwidth]{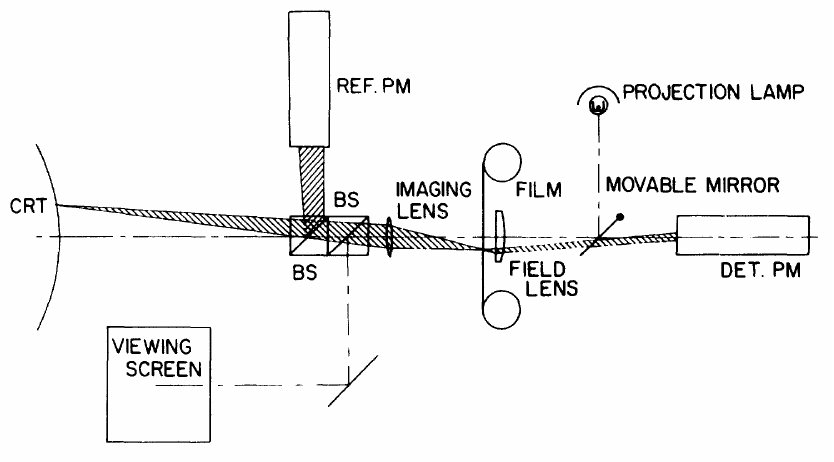}
        \caption{A spot scanner using CRT as a moving light source.
        			The light spot was controlled by a PDP-1 computer, and was focused onto the film,
			of which the local transparency was measured by a photomultiplier (DET. PM).
			{\footnotesize @2014 IEEE. Reprinted, with permission, from}
        		\cite{wenzel_ieee_13_1966}.}
        \label{fig:wenzel_ieee_13_34_1966_fig5}
\end{wrapfigure}
In 1969, %to scan a larger number of photographs, 
the experiment at the CERN PS~\cite{bohm_npb9_1969}
used an automatic ``Luciole Flying Spot Digitizer''
which could scan 1000 frames per hour.
This used a CRT instead of a mechanical system
to move a bright light spot quickly across a film
and record the intensity of light passing through the film with a phototube~\cite{wenzel_ieee_13_1966}.
Figure~\ref{fig:wenzel_ieee_13_34_1966_fig5} shows a similar spot scanner with a CRT,
 made by the Univ. of Michigan.

In the late 1960s, experiments started to move away from using films.
Experiments at CERN and BNL used ferrite-core readout systems
to read out spark positions and recorded them on tape directly~\cite{faissner_pl30b_1969, jensen_prl23_1969, fryberger_ieee15_1968}.
This readout system allowed 
the BNL experiment to collect 9400 $K_{L, S} \to \pi^+\pi^-$ events, 300 times the first \CP\  violation experiment.

In the early 1970s, experiments started to use multi-wire proportional chambers (MWPC).
For example, the experiment that observed the first 
$K_L \to \mu^+\mu^-$ decays~\cite{carithers_prl30_1336_1973}
used MWPCs with 5000 wires that had a 2-mm wire spacing.
With the same detector, the experiment collected 2 M 
\(K_{L, S} \to \pi^+\pi^-\) events~\cite{carithers_prl34_1244_1975}.

The geometry of spectrometers has also changed.
The \CP\  violation experiment in 1964 used a double-arm spectrometer with two sets of quadrupole 
magnets and spark chambers located at \(\pm22^\circ\) from the neutral beam line
as shown in Fig.~\ref{fig:christensen_prb140_1965_Fig1}.
This geometry was optimized for the 1.1-GeV/c average $K_L$ momentum.

\begin{figure}[!ht]
    \begin{center}
        \includegraphics[width=0.8\columnwidth]{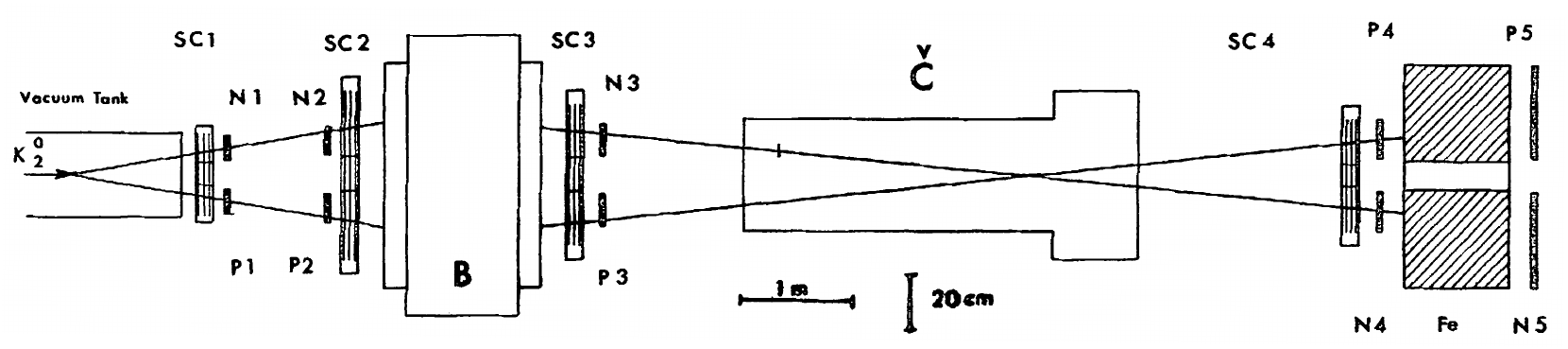}
        \caption{Plan view of an experimental apparatus at CERN which used a forward spectrometer.
        			SC$i$ are spark chambers, 
        			N$i$ and P$i$ are scintillators, B is a bending magnet, and \v{C} is a \v{C}erenkov counter.
        		{\footnotesize Reprinted from}~\cite{bouard_pl15_1965}.
		{\footnotesize Copyright 2014, with permission from Elsevier.}
		}
        \label{fig:DeBouard_pl15_58_1965_1965_fig1}
    \end{center}
\end{figure}

A later experiment at CERN~\cite{bouard_pl15_1965} in 1965 used a forward spectrometer
with one dipole magnet sandwiched between four spark chambers
as shown in Fig.~\ref{fig:DeBouard_pl15_58_1965_1965_fig1}
to have a high acceptance for pions from $K_L$'s with the average momentum 11 GeV/c.
It also had a \v{C}erenkov counter between the downstream spark chambers 
for a particle identification.
The forward spectrometer became the standard for the experiments that followed.

%-------------------------------------------------------
\subsubsection{Photon Detectors}
Let us now switch to the neutral $\klpiopio$ mode.
One of the questions after the discovery of \CP\  violation was 
whether there was also the neutral counterpart of the $\klpipi$.
The $\klpiopio$ decay mode was far more difficult than the charged mode,
because photons were not readily visible, and 
there was a large background from the $\klpiopiopio$ decay mode.

In 1968, an experiment at CERN~\cite{bugadov_plb28_1968} used a 
bubble chamber, 1.2 m in diameter and 1 m deep, filled with heavy liquid freon
to collect events with 4 photons converted in the chamber.
The experiment observed 24 events with 7.4 background events, and gave 
\(BR(\klpiopio) = (0.94 \pm 0.37) \times 10^{-3}\).

\begin{figure}[!ht]
	\centering
	\begin{minipage}[t]{0.48\columnwidth}
		\includegraphics[width=1.0\columnwidth]{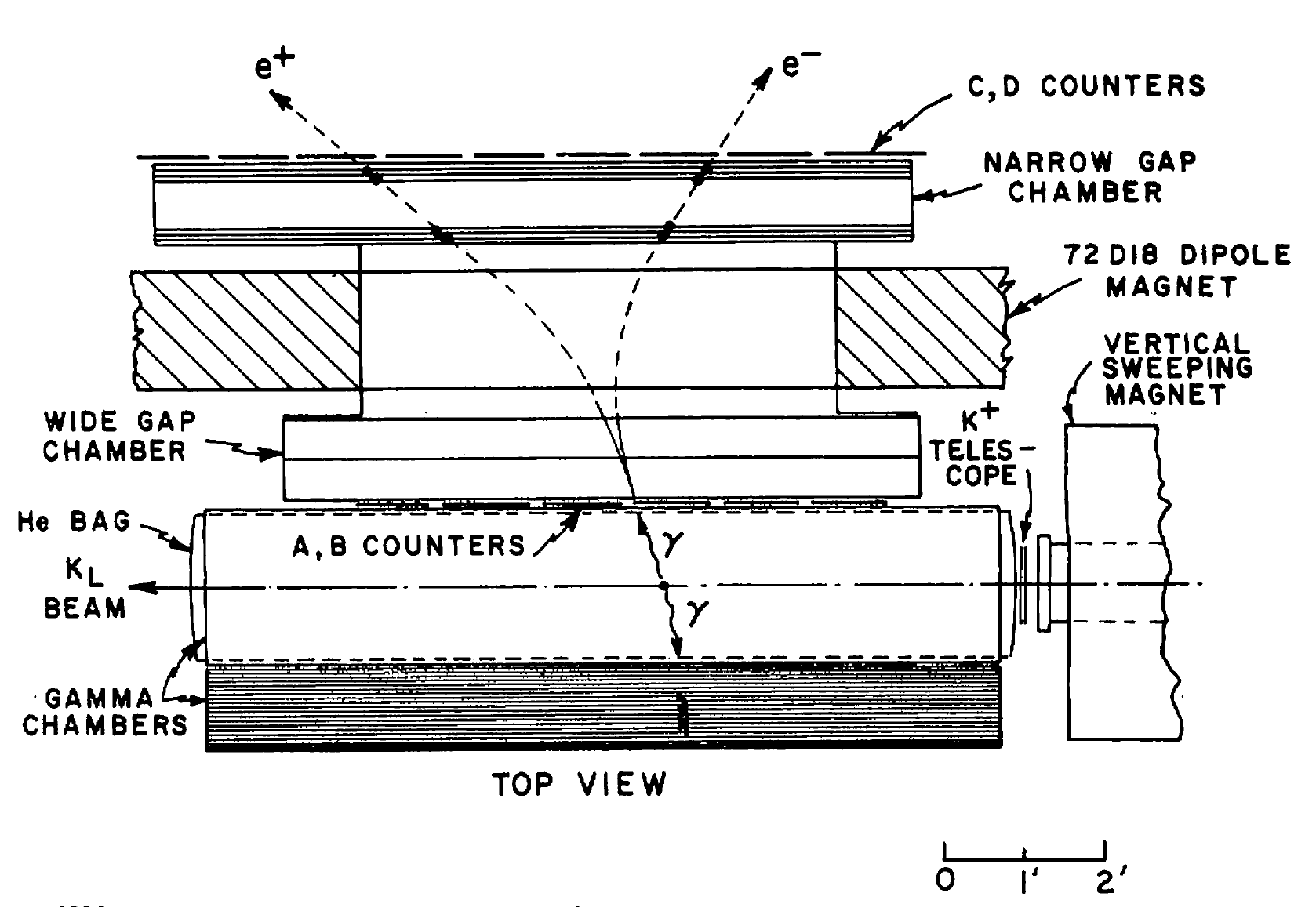}
		\caption{Plan view of the experimental apparatus that observed $\klpiopio$ events.
				The $K_L$ beam entered from the right, and photon detectors surrounded 
				the \emph{sides} of the beam~\cite{banner_prl21_1103_1968}.
				\copyrightAPS
				}
    		\label{fig:banner_prl21_1103_1968_Fig1a}
	\end{minipage}
	\hspace{0.03\columnwidth}
	\begin{minipage}[t]{0.45\columnwidth}
                \includegraphics[width=\columnwidth]{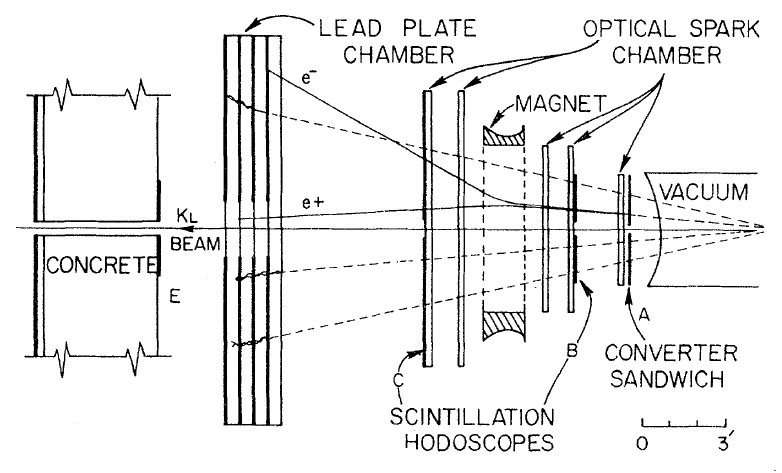}
                \caption{Schematic view of the experimental apparatus at BNL
                		that was used to detect $\klpiopio$ decays.
			Photon detectors were placed in the beam direction
			~\cite{banner_prl28_1972}.
			\copyrightAPS
			}
                        \label{fig:banner_prl28_1597_1972_Fig1}
	\end{minipage}
\end{figure}

Also in 1968, an experiment at Princeton-Pennsylvania Accelerator~\cite{banner_prl21_1107_1968, banner_pr188_1969}
used a spectrometer and ``$\gamma$ chambers'' 
surrounding the four sides of a 30 cm $\times$ 30 cm $K_L$ beam 
as shown in 
Fig.~\ref{fig:banner_prl21_1103_1968_Fig1a}.
A photon was converted in a 0.1 $X_0$ lead sheet placed on one side, 
and the resulting electron pair was momentum analyzed in a magnetic spectrometer.
A wide gap spark chamber was used before the magnet to minimize multiple scatterings,
and thin gap spark chambers were used to track the $e^\pm$ pairs after leaving the magnet.
The other three sides were covered with ``$\gamma$ chambers'' 
consisting of layers of steel plates and spark chambers to 
measure the hit positions of the other three photons.
The hit timings were also recorded to measure the time-of-flight of 
$K_L$s with a peak momentum of 0.25 GeV/c  bunched in a 1 ns width.
The decay vertex was assumed to be along the photon momentum axis 
measured with the spectrometer to reconstruct the decay.
The experiment observed 57 $\pm$ 9 events, and gave
\(BR(\klpiopio) = (0.97 \pm 0.23) \times 10^{-3}\).

The same group then moved to BNL to measure 
\(|\eta_{00}|/|\eta_\pm|\).
The detector geometry was changed 
to have a higher acceptance for a higher mean $K_L$ momentum (6 GeV/c), 
and to detect both $\pi^+\pi^-$ and $\pi^0\pi^0$ modes.
As shown in Fig.~\ref{fig:banner_prl28_1597_1972_Fig1}, 
a spectrometer and ``$\gamma$ counters'' were located in the
beam-forward direction.
Still it required one of the four photons to be converted in a 0.1-$X_0$-thick converter, 
and the momenta of converted electron pairs to be measured.
The $\pi^0\pi^0$ events were reconstructed basically with the same technique
as the previous experiment,
with the momentum of one photon, along with conversion positions, but not the energies,
of other three photons.
The experiment observed 124 $\pm$ 11 $\klpiopio$ events 
with 3 $\pm$ 3 $\klpiopiopio$ background events~\cite{banner_prl28_1972}.

\subsubsection{Calorimeters}
\begin{figure}[!ht]
	\centering
	\begin{minipage}[t]{0.42\columnwidth}
            \includegraphics[width=\columnwidth]{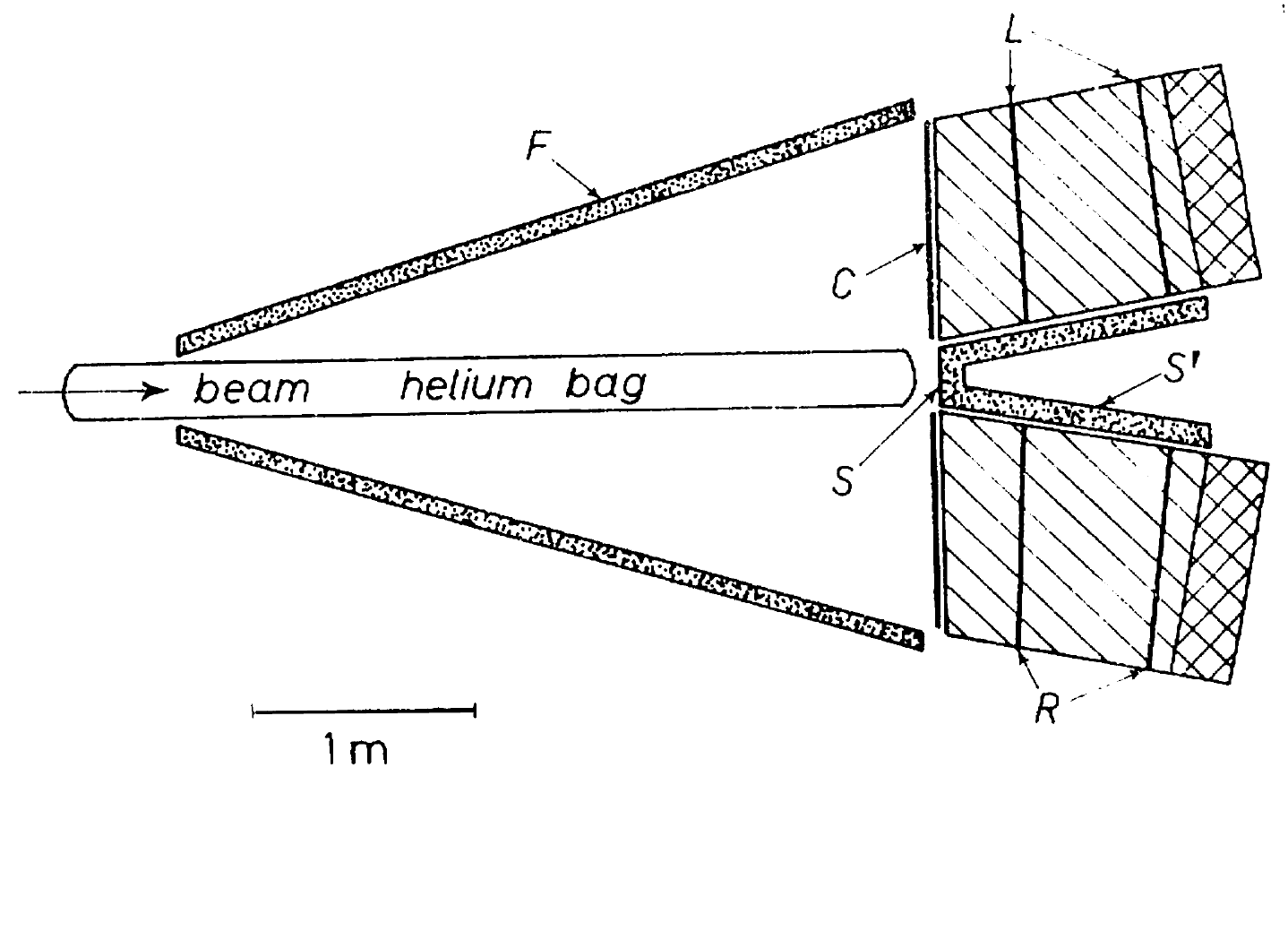}
            \caption{Plan view of the experiment apparatus at CERN (taken from~\cite{gaillard_nuovo_59a_1969}).
            		Two spark chambers systems with radiators were used to measure conversion points, 
    		initial direction, and shower energy of each photon from $\klpiopio$.
		{\footnotesize With kind permission of Springer Science+Business Media.}
		}
            \label{fig:gaillard_nuovo_59a_453_1969_Fig2}
	\end{minipage}
	\hspace{0.01\columnwidth}
	\begin{minipage}[t]{0.5\columnwidth}
            \includegraphics[width=\columnwidth]{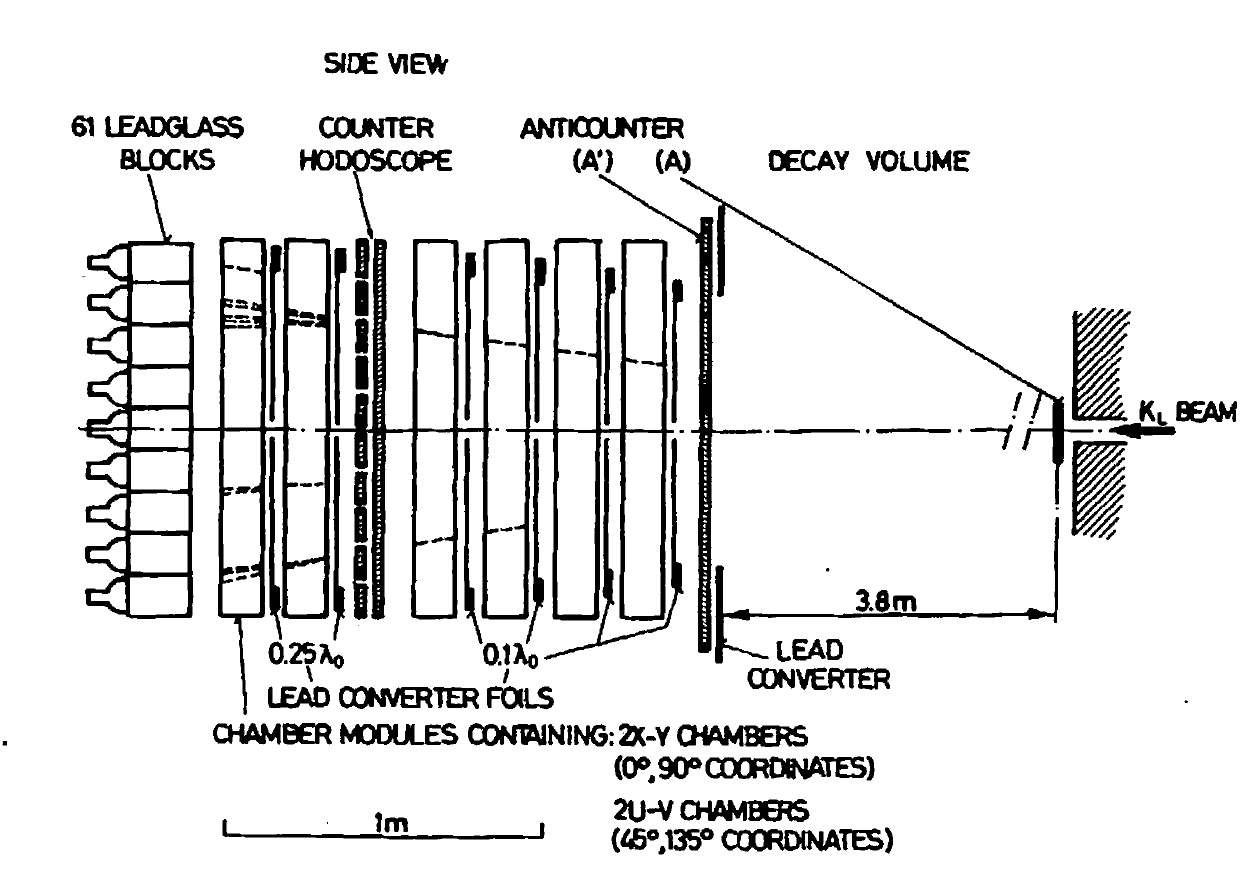}
            \caption{Plan view of the experiment apparatus at CERN 
            		using a lead glass array to measure the energy of 
    		photons from $\klpiopio$.
            		Spark chambers interleaved with converter foils were used to 
    		measure the conversion points and directions of two photons from the decay.
		{\footnotesize Reprinted from}~\cite{holder_plb40_1972}.
		{\footnotesize Copyright 2014, with permission from Elsevier.}
		}
            \label{fig:holder_plb40_141_1972_Fig1}
	\end{minipage}
\end{figure}

There was also an attempt to measure the
energies and directions of \emph{all} four photons with limited accuracies.
An experiment at CERN~\cite{gaillard_nuovo_59a_1969} 
used two spark chamber systems with Al and brass plates, with a total thickness of 
11.6 $X_0$, as shown in Fig.~\ref{fig:gaillard_nuovo_59a_453_1969_Fig2}.
The number of sparks gave an energy resolution of 25\% for 500 MeV electrons.
The experiment observed about 200 $\klpiopio$ events, and gave
\(BR(\klpiopio) = (2.5 \pm 0.8) \times 10^{-3}\).

Another experiment at CERN~\cite{holder_plb40_1972} introduced
a calorimeter consisting of 61 hexagonal lead-glass modules 
(13 $X_0$ long) to measure the energies of \emph{all four} photons,
as shown in Fig.~\ref{fig:holder_plb40_141_1972_Fig1}.
The energy resolution was 3.3\%/$\sqrt{E_\gamma \mathrm{(GeV)}}$.
The experiment also had spark chambers with lead foils to 
measure the directions of at least two photons to reconstruct 
the decay vertex.
The pulse heights from the lead-glass and spark coordinates read out via ferrite cores were
written on tape.
The experiment collected 167 $\klpiopio$ events, and gave
\(|\eta_{00}/\eta_\pm| = 1.00 \pm 0.06\).

%-------------------------------------------------------
\subsection{Hard Wall}

By 1973, the number of $\klpipi$ events reached 4200~\cite{messner_prl30_1973}.
However, the number of $\klpiopio$ events was still about 200.
The small statistics of $\klpiopio$ events limited the accuracy of $\reepoe$;
the best results were
\(\reepoe = -0.016 \pm 0.040\)~\cite{banner_prl28_1972}, and
\(\reepoe = 0.00 \pm 0.01\)~\cite{holder_plb40_1972}, 
which were both consistent with 0. 
Experiments hit a hard wall.

%% Banner 1972: |eta00/eta+- |^2 = 1.05 +- 0.14
%	e'/e = -0.0155 +- 0.0403
%	
%% Holder: |eta00 / eta+-| = 1.00 +- 0.06
%	e'/e = 0 +- 0.01

In 1976, in his beautiful review, Kleinknecht wrote~\cite{kleinknecht_1976}
\begin{quote}
	It is not easy to improve substantially the experimental precision.
	A decision between superweak and milliweak models of \CP\ violation
	will therefore probably have to come from other experimental information
	outside the $K^0$ system.
\end{quote}
Considering the difficulties that they had in the mid-1970s, 
such a view is not surprising.

What did kaon experiments do then?
There were two major streams.

One was to measure the charge asymmetry of semi-leptonic 
$K_L \to \pi e \nu$ and $K_L \to \pi \mu \nu$ decays.
The charge asymmetry, 
$\delta = (N^+ - N^-) / (N^+ + N^-)$, gives $2Re(\epsilon)$
where $N^\pm$ is the number of observed $K_{\ell 3}\) decays with $\ell^\pm$.
These measurements required high statistics and a good control of systematic errors.
For example, an experiment at CERN~\cite{geweniger_plb48_1974}
collected 34M $K_{e3}$ and 15M $K_{\mu 3}$ events, and measured 
\(\delta_e = (3.65 \pm 0.17)\times 10^{-3}\) and
\(\delta_\mu = (3.23 \pm 0.26) \times 10^{-3}\).

The other stream was to measure the amplitudes of 
the regeneration of $K_S$ from $K_L$ interacting with materials.
Although it looked irrelevant, 
this stream lead to the measurements of $\reepoe$ later.

%-------------------------------------------------------

\subsection{Standard Model Prediction on the $\epoe$}
\begin{figure}[!ht] %  figure placement: here, top, bottom, or page
	   \centering
	   \subfigure[]{
		\includegraphics[width=0.3\linewidth]{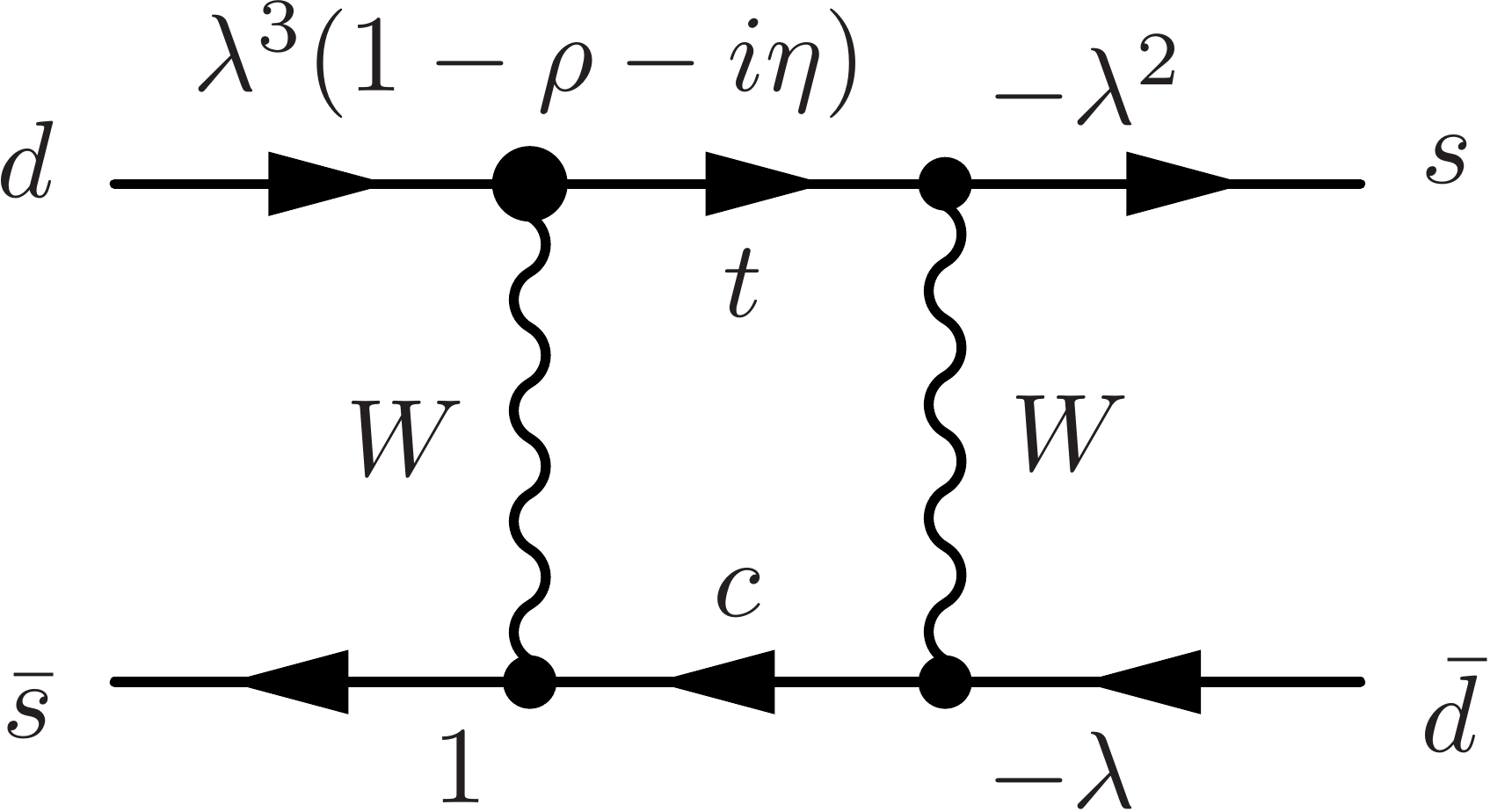}
    		}
    		\hspace{5mm}
	   \subfigure[]{
		\includegraphics[width=0.24\linewidth]{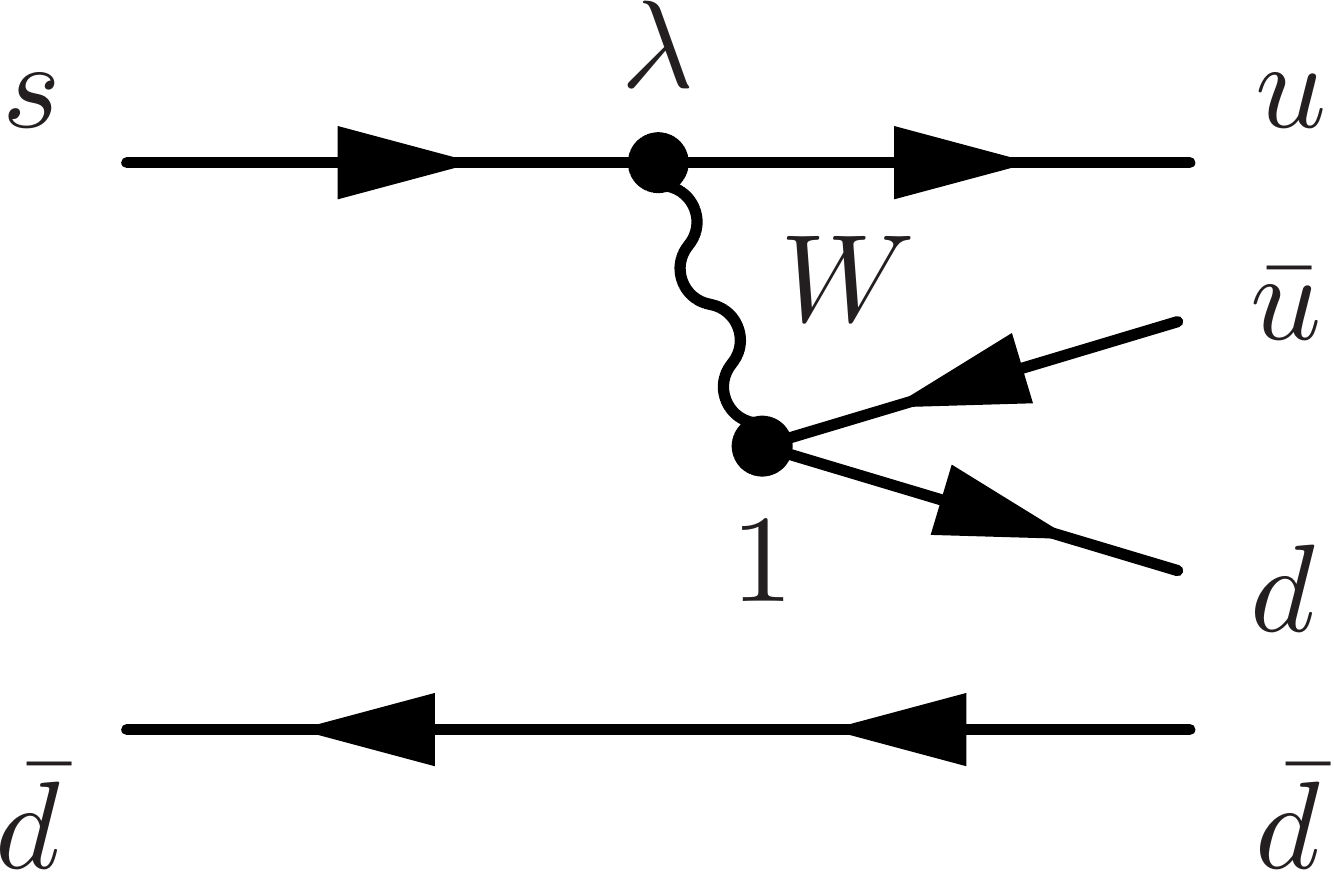}
    		}
		\hspace{5mm}
	   \subfigure[]{
		\includegraphics[width=0.27\linewidth]{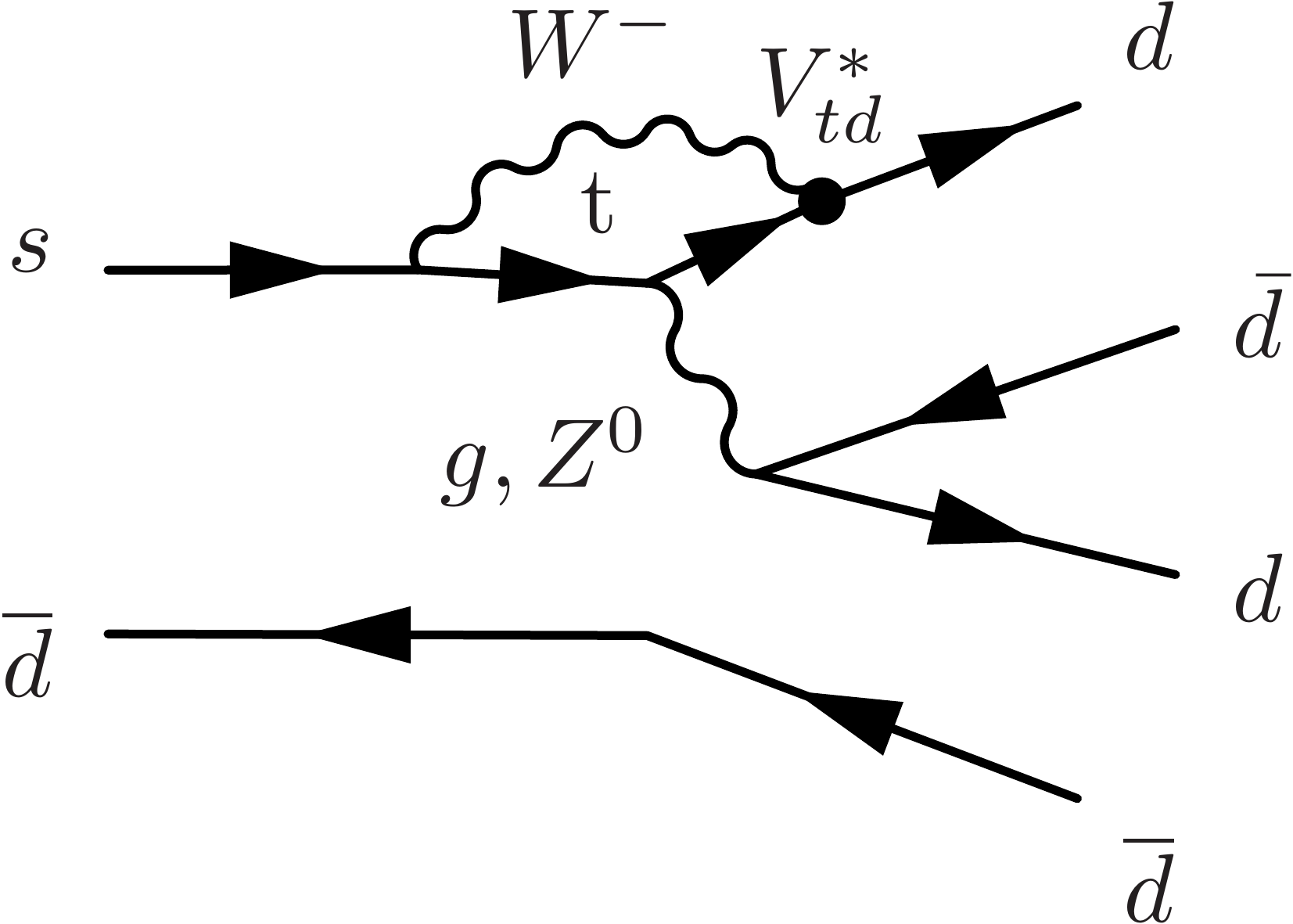}
    		}
	   \caption{(a):Box diagram that introduces indirect \CP\  violation in $\ko - \kobar$ mixing.
	   		(b): Tree diagram for \(K \to \pi\pi\).
			(c): Penguin diagram that can generate direct \CP\  violation.}
	   \label{fig:k2pidiagrams}
\end{figure}

In the 1970s, there were movements on the theoretical side.
First, Kobayashi and Maskawa explained~\cite{km_1974} that the phase 
which naturally comes in the mixing between 
3 generations of quarks can explain the \CP\  violation in the 
\(K^0 - \overline{K^0}\) transition via a box diagram
shown in Fig.~\ref{fig:k2pidiagrams}(a).
Second, 
in addition to a standard tree diagram shown in Fig.~\ref{fig:k2pidiagrams}(b),
a so-called penguin diagram for the $\klpipi$ decay 
shown in Fig.~\ref{fig:k2pidiagrams}(c)
was introduced by 
Ellis \etal~\cite{ellis} and 
Shifman \etal~\cite{vainshtein_jetplett_22_1975, shifman_npb_120_1977}.
The complex phase in the penguin diagram can violate \CP\  in the decay process.
Gilman and Wise predicted the value of $\reepoe$ 
to be \((3 - 5) \times 10^{-3}\) with the 
Kobayashi-Maskawa model~\cite{gilman_prd20_1979},
whereas the superweak model predicted it to be 0.
To measure the $\reepoe$ with an accuracy of \(1 \times 10^{-3}\), 
30k \(\klpiopio\) events are needed, 
and systematic errors should be controlled to a 0.1\% level.
Although the predicted value was 10 times smaller than
the values predicted by some other models, 
it gave a clear goal for experiments.

\subsection{Precision Experiments}

\subsubsection{Early 1980s}
\begin{figure}[!ht] %  figure placement: here, top, bottom, or page
	   \centering
	   \subfigure[]{
		\includegraphics[width=0.35\linewidth]{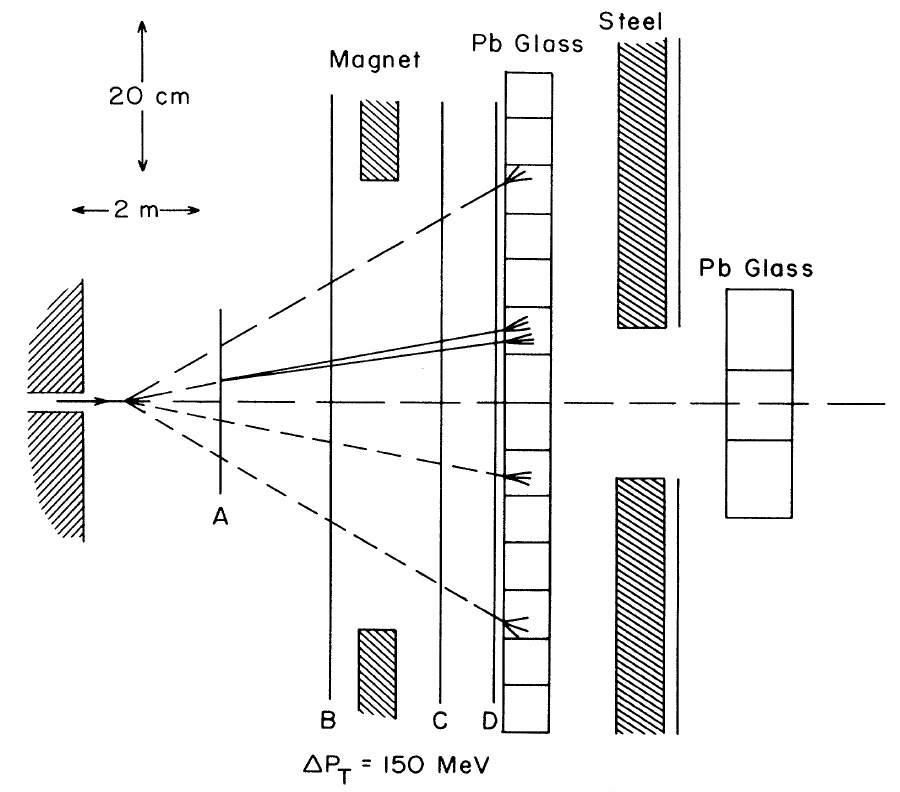}
    		}
    		\hspace{5mm}
	   \subfigure[]{
		\includegraphics[width=0.55\linewidth]{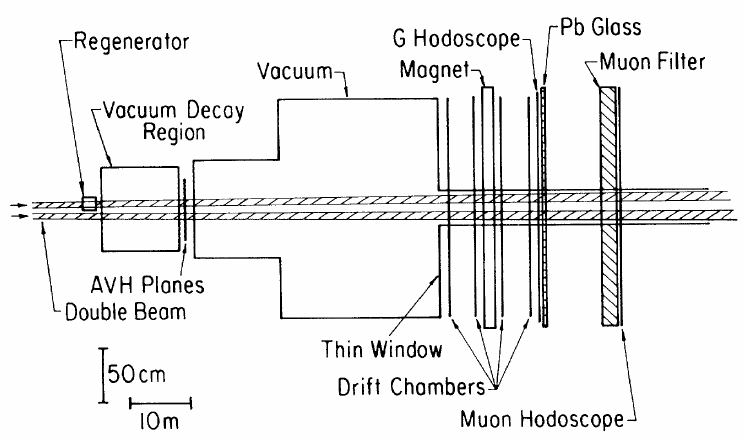}
    		}
	   \caption{(a): Schematics of the $\reepoe$ experimental apparatuses at BNL (taken from \cite{black_prl54_1985}, 
	   and (b): Fermilab E617 (taken from \cite{bernstein_prl54_1985}).
	   Fermilab E617 used two beams to observe $K_L$ and $K_S$ decays simultaneously.
	   \copyrightAPS
	   }
	   \label{fig:black_bernstein}
\end{figure}

In the early 1980s, a new generation of experiments started
at BNL~\cite{black_prl54_1985} 
and Fermilab (FNAL E617)~\cite{bernstein_prl54_1985}.
Both experiments used dipole magnets and chambers 
(MWPCs for BNL, and drift chambers for Fermilab)
to analyze the momentum of 
$\pi^+\pi^-$ tracks,
as shown in Fig.~\ref{fig:black_bernstein}.
A thin lead sheet was placed upstream of the first chamber
to convert one of the photons from the $\pi^0\pi^0$ decays, 
and to track the produced electron pairs.
They both used lead glass arrays to measure the hit positions and energies of 
photons and electrons. 
The \(K_L \to \pi e \nu\) events were rejected by comparing the energy deposit in the 
lead glass array and the track momentum.
The \(K_L \to \pi \mu \nu\) events were rejected by detecting muons passing through a steel wall
located downstream of the calorimeter.
Both experiments inserted carbon blocks in the $K_L$ beam to regenerate $K_S$, 
but the techniques were different.
In the BNL experiment, the regenerator was moved in and out to alternate between $K_S$ and $K_L$ runs.
The Fermilab experiment had two $K_L$ beams, and moved the regenerator between 
the two beams
to observe $K_L$ and $K_S$ decays simultaneously, which cancels various systematical effects.
Another difference was the proton and kaon momentum;
the BNL experiment used 28-GeV protons to produce 7-14 GeV/c kaons, while
the Fermilab experiment used 400-GeV protons to produce kaons with the mean momentum around 70--90 GeV/c.
The BNL experiment collected 1120 $\klpiopio$ events and gave
\(\reepoe = 0.0017 \pm 0.0082\).
The Fermilab experiment collected 3150 $\klpiopio$ events and gave 
\(\reepoe = -0.0046 \pm 0.0058\).
Although both results were consistent with 0, 
it became clear that using higher energy kaons was more advantageous
because of the higher acceptance and the better energy resolution for photons.

It is worth noting that the Fermilab E617 was the cornerstone for the Fermilab $\epoe$ experiments
that followed.
The experiment introduced a double beam technique, using a regenerator
%\footnote{There was even an experiment which inserted regenerators in parts of one beam to 
%make three beam cross sections with different regenerator configurations\cite{cullen_plb32_1970} 
%to measure 
%the $K_L - K_S$ mass difference.}
to observe $K_L$ and $K_S$ decays simultaneously to suppress systematic errors.
This double-beam technique was actually inherited from 
Fermilab E226 and E486 which studied regeneration on electrons~\cite{molzon_prl41_1978} and
coherent regeneration amplitudes on various nuclei~\cite{gsponer_prl42_13_1979}, respectively.

\subsubsection{Fermilab E731}
\begin{figure}[!ht]
    \begin{center}
        \includegraphics[width=0.7\columnwidth]{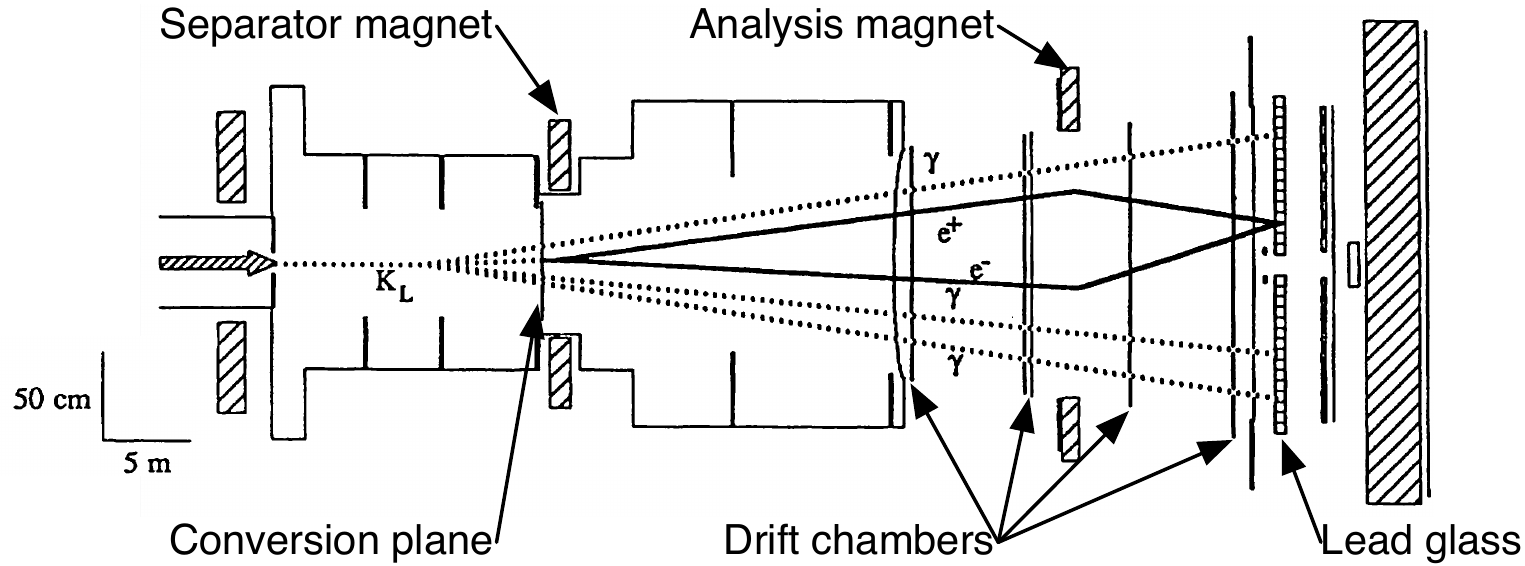}
        \caption{Plan view of the Fermilab E731 apparatus. 
        		Two beams were aligned vertically.
		Initially, for the $\klpiopio$ decay mode, one of the photons were required to be converted.
        		}
        \label{fig:e731_conversion}
    \end{center}
\end{figure}

In the mid-1980s, Fermilab E731 was built.
It used 800 GeV protons to produce two $K_L$ beams with the average energy of 70 GeV.
An improved \ce{B4C} regenerator alternated between the two beams every spill.
Four drift chambers were built for a better position resolution, and 
a lead glass array was used as an electromagnetic calorimeter.
Initially, for the $\pi^0\pi^0$ run, a 0.1-$X_0$-thick lead sheet was inserted at the end of 
a decay volume to convert one of the photons for tracking, just as the past experiments.
A dipole magnet located just downstream of the lead sheet split the electron pairs.
Its field was tuned such that 
the downstream spectrometer would focus the pair on to the lead glass calorimeter, 
as shown in Fig.~\ref{fig:e731_conversion}.
However, requiring a photon conversion imposed a limit on the statistics;
only 30\% of the $\pi^0\pi^0$ decays had converted electron pairs, 
and the $\pi^+\pi^-$ and $\pi^0\pi^0$ events had to be collected in separate runs
because the lead sheet had to be removed for the $\pi^+\pi^-$ run to minimize multiple scatterings.

After some studies, the experiment decided to break the 30 years of tradition of converting photons.
For the last part of the experiment, it ran without the lead sheet, and collected 4 photons without converting them,%
\footnote{CERN NA31 experiment triggered on $2\pi^0$ events with a hodoscope in a liquid Ar calorimeter.} 
and also took $\pi^+\pi^-$ and $\pi^0\pi^0$ modes simultaneously for the first time.
This big step was made possible by two technological developments.
One was a second-level trigger to count the number of clusters in the calorimeter~\cite{hcf}
which reduced the trigger rate by a factor 10 by collecting only 4 and 6 clusters.
The other was a FASTBUS-based data acquisition system which reduced the dead time 
by a factor 10, from 5 ms to 0.5 ms.

At the end, Fermilab E731 collected 410 k $\klpiopio$ events and 327 k $\klpipi$ events, 
and gave
\(\reepoe = (7.4 \pm 5.2 \mathrm{(stat)} \pm 2.9 \mathrm{(syst)}) \times 10^{-4}\)~\cite{e731final}.
Compared to E617, it reduced the overall error by 
a factor 14 with 130 times more $\klpiopio$ events.

The Fermilab result was only 1.2$\sigma$ away from 0, and still consistent with 0,
whereas its competitor, CERN NA31, gave 
\(\reepoe = (20 \pm 7) \times 10^{-4}\)~\cite{barr_plb_317_1993} which was 3$\sigma$ away from 0.
There were intense arguments between the two experiments.
CERN NA31 used a moving production target for $K_S$ runs
to make the $K_L$ and $K_S$ decay vertex distributions similar.
The experiment claimed that this method made the analysis less dependent on 
Monte Carlo simulations.
%Also, by not using a magnetic spectrometer but instead using a hadronic calorimeter to
%measure $\pi^+\pi^-$ energies and positions, they claimed to have similar acceptances
%between charged and neutral modes.
Fermilab E731 argued that even with different decay vertex distributions between $K_L$ and 
regenerated $K_S$, the geometrical acceptances obtained with Monte Carlo simulations 
can be checked with high-statistics decay modes
such as $K_L \to \pi e \nu$ and \(\klpiopiopio\).
Also, it argued that it is more important to collect $K_L$ and $K_S$ data simultaneously 
with the same detector, 
rather than to take $\pi^+\pi^-$ and $\pi^0\pi^0$ modes simultaneously, 
because efficiencies of charged and neutral mode detectors have different rate dependences.
These arguments, however, could not solve the discrepancy between the two experimental results.
At the end, both groups decided to build new experiments,
CERN NA48, and Fermilab KTeV-E832,
to improve the accuracy and precision by another order of magnitude.
In this paper, I will concentrate on KTeV-E832, because CERN NA48 will be covered in detail
by M.~Sozzi~\cite{sozzi}.

\subsubsection{Fermilab KTeV-E832}
\begin{figure}[!ht]
	\centering
	\subfigure[]{\includegraphics[width=0.6\columnwidth]{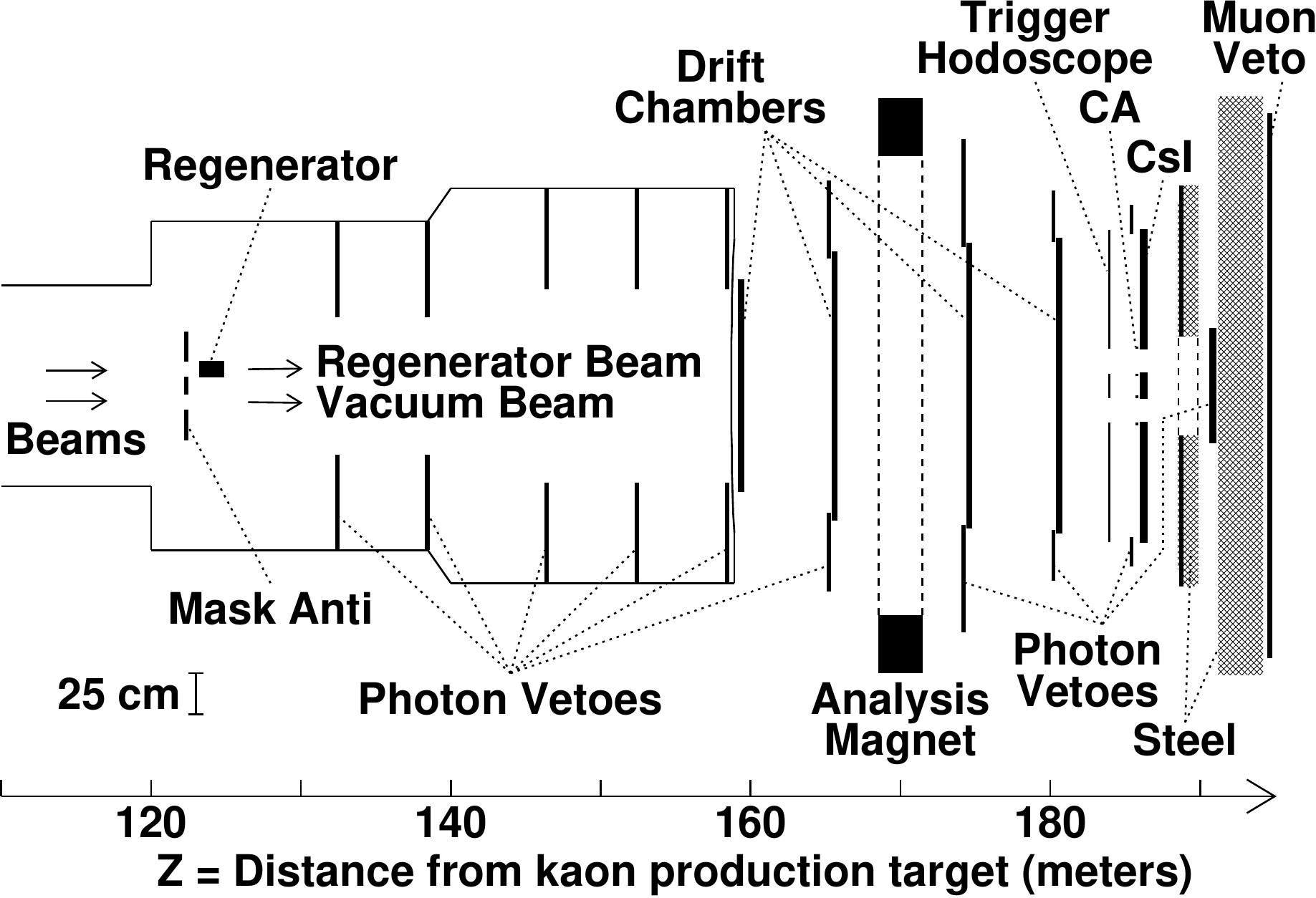}}
	\hspace{5mm}
	\subfigure[]{\includegraphics[width=0.275\linewidth]{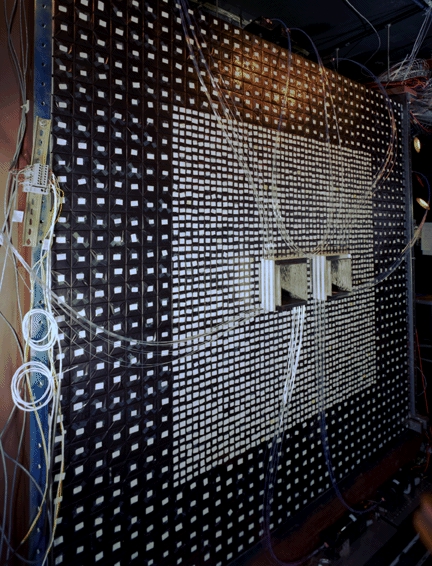}}
	\caption{(a) Plan view of the Fermilab KTeV apparatus
        		(taken from \cite{e832_final_2011}).
		\copyrightAPS\ 
			(b) KTeV CsI electromagnetic calorimeter.
	}
	\label{fig:ktev_det_csi}
\end{figure}

Fermilab KTeV built a new beam line with a better collimation scheme to 
make a cleaner beam with less halo neutrons even with a higher proton intensity.
Figure~\ref{fig:ktev_det_csi}(a) shows the plan view of the KTeV detectors.
The new regenerator was made of fully active scintillators to suppress backgrounds from
 non-coherent regenerations.
A new large spectrometer magnet with a uniform field was made for the $\pi^+\pi^-$ decays.
A new electromagnetic calorimeter shown in Fig.~\ref{fig:ktev_det_csi}(b), covering 1.9 m $\times$ 1.9 m,
 was built with pure CsI crystals 27 $X_0$ long, 
to have a better energy resolution.
To improve the position resolution, 
2.5-cm-square crystals were used in the central 1.2 m $\times$ 1.2 m region,
and 5.0-cm-square crystals in the outer region.
The scintillation light from the crystals were read out by phototubes
and digitized right behind the phototubes every 19 ns
to minimize electric noise, record the pulse shape, 
and to supply hit information for counting the number of clusters.
%As shown in Fig.~\ref{fig:ktev_csi}(b), 
The energy resolution was less than 1\% for most of the energy range, reaching about 0.45\%, 
which was the best resolution for the calorimeter of this size.
A new data acquisition system used a matrix of memories
to buffer events during spill%
\footnote{%
The dead time of the trigger and data acquisition system was only 10 $\mu$s, 1/50 of that of E731.
It could accept 8 kbyte events coming at 20 kHz for 20 s spill with a 60 s cycle.}
and to distribute the events evenly to 36 120-MIPS CPUs (fast in those days)
to reconstruct all the events for online filtering.

\begin{wrapfigure}{r}{0.52\columnwidth}
        \includegraphics[width=0.52\columnwidth]{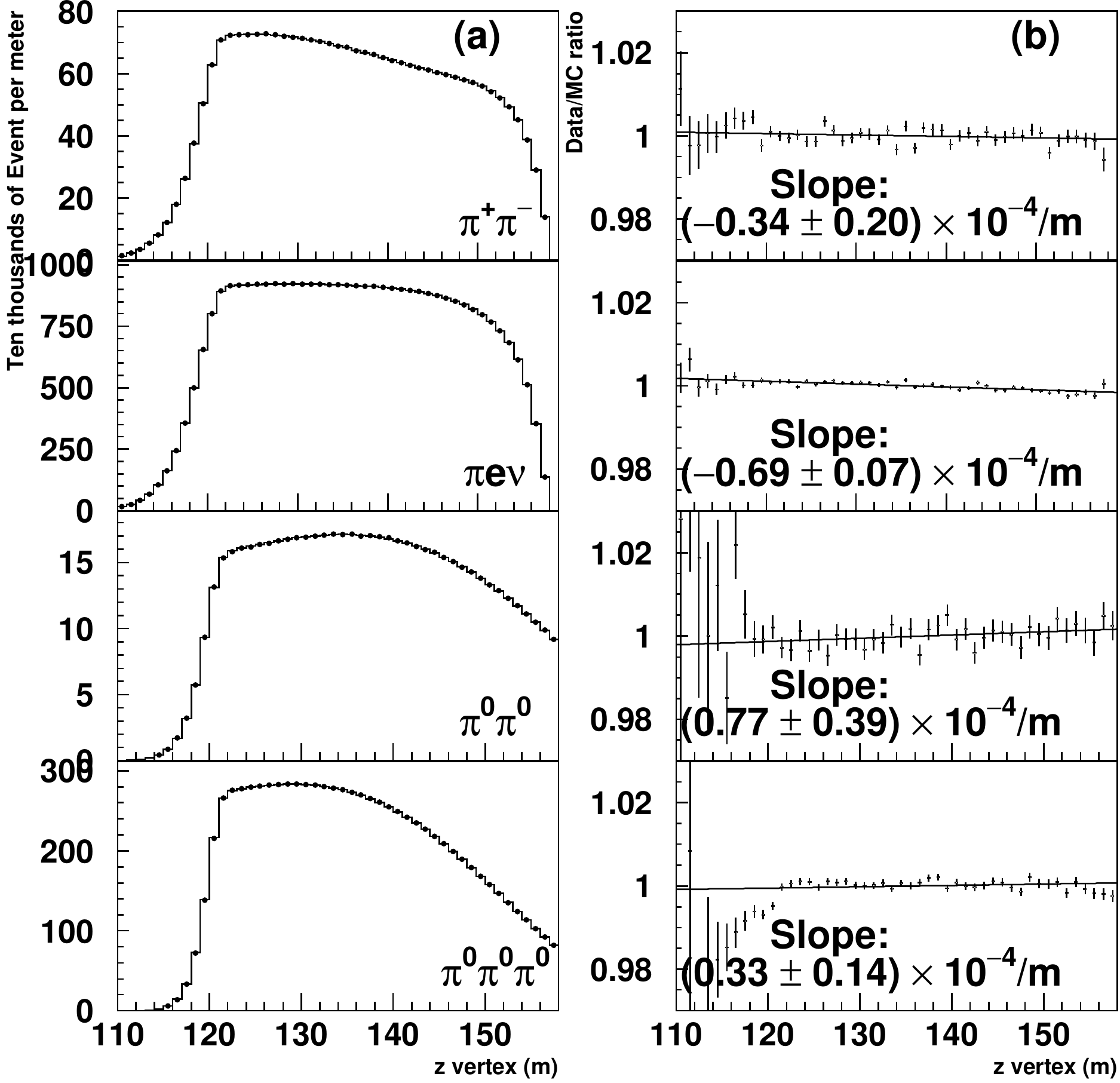}
        \caption{(a) Decay vertex distributions of
        			\(K_L \to \pi^+\pi^-\), \(K_L \to \pi e\nu\), \(K_L \to \pi^0\pi^0\), and 
			\(K_L \to \pi^0\pi^0\pi^0\) decays
			for the data (dots) and MC (histogram).
        			(b) The data-to-MC ratios are fit to a line
        		(taken from \cite{e832_final_2011}).
		\copyrightAPS
        		}
        \label{fig:e832_vtxz}
\end{wrapfigure}

Figure \ref{fig:e832_vtxz} shows the decay vertex distributions for the $\pi^+\pi^-$, 
$\pi^0\pi^0$, and other high statistics decay modes.
The number of $\klpiopio$ events was 6M.
The data and Monte Carlo 
distributions agreed well, and the systematic errors on the $\reepoe$ due to acceptance correction were 
\(0.57 \times 10^{-4}\) for charged modes and 
\(0.48 \times 10^{-4}\) for neutral modes.
The first result based on 20\% of data taken in 1996-1997 runs was 
\(\reepoe = (28.0 \pm 4.1) \times 10^{-4}\)~\cite{e832_prl83_1999}, 
7 $\sigma$ away from 0, and 
the final result with the full data was 
\(\reepoe = (19.2 \pm 1.1 \mathrm{(stat)} \pm 1.8 \mathrm{(syst)}) \times 10^{-4}
 = (19.2 \pm 2.1) \times 10^{-4}\)~\cite{e832_final_2011}.

\subsection{Conclusion on the $\epoe$}
 CERN NA48 gave \(\reepoe = (14.7 \pm 2.2) \times 10^{-4}\) based on all data~\cite{cern_final_2002}.
 The result averaged by PDG is \((16.6 \pm 2.3) \times 10^{-4}\)~\cite{pdg_epoe}, 
7.2$\sigma$ away from 0.
 CERN NA48 and Fermilab KTeV-E832 have both clearly established 
 that the $\reepoe$ is not 0, thereby rejecting the Superweak model, 
 and supported the Kobayashi-Maskawa model.
 \clearpage
 
\subsection{Looking Back}
\begin{figure}[!ht]
	\centering
        \includegraphics[width=\columnwidth]{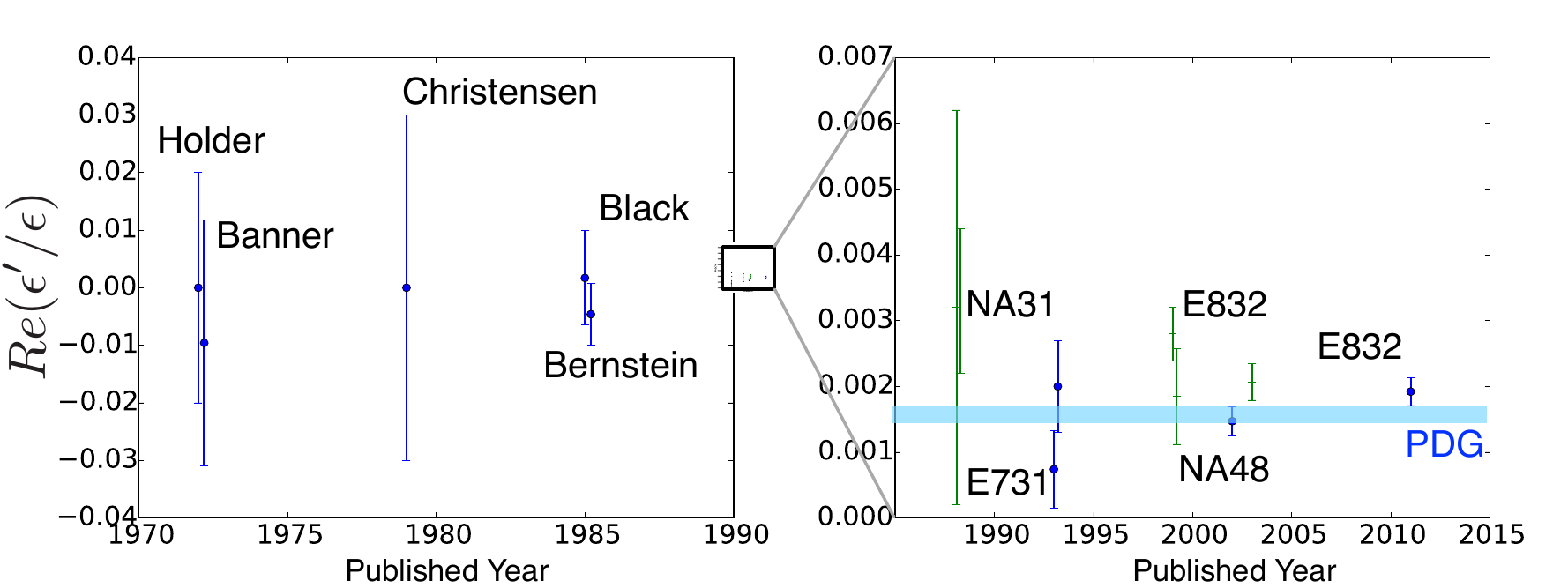}
        \caption{The results of $\reepoe$ vs. published year.
        	The horizontal band shows the PDG average \cite{pdg_epoe}.}
        \label{fig:epoe_history}
\end{figure}

\begin{wrapfigure}{r}{0.5\columnwidth}
        \includegraphics[width=0.5\columnwidth]{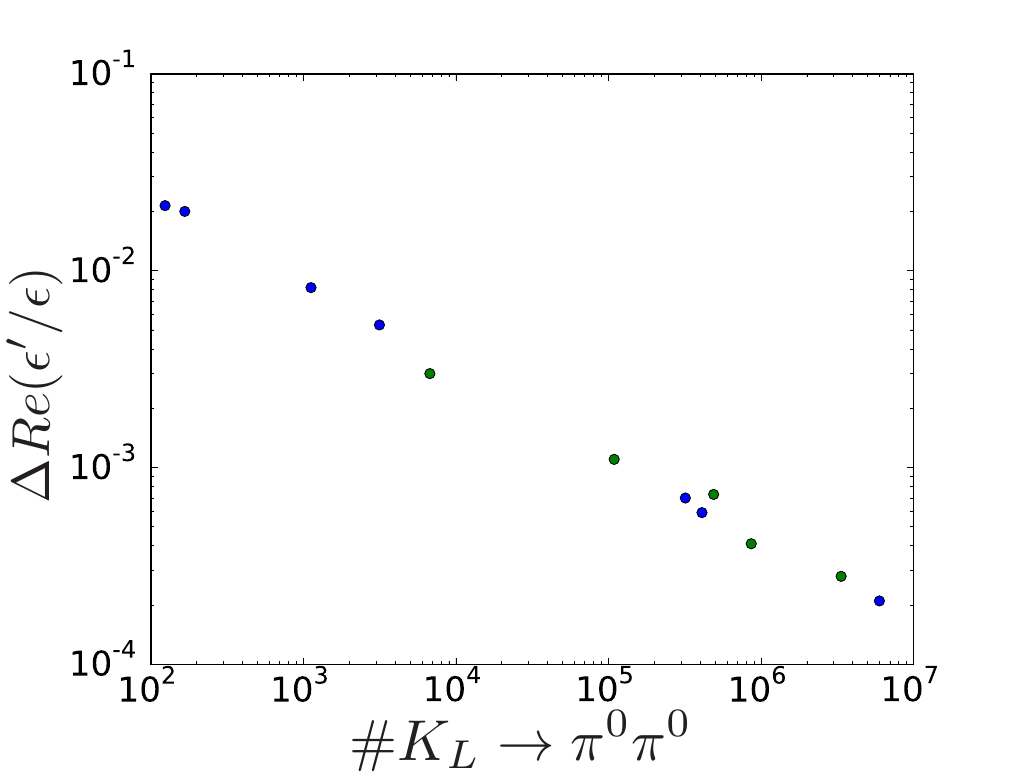}
        \caption{The errors on $\reepoe$ are shown as a function of the number of $\klpiopio$ events for the past experiments.}
        \label{fig:kl2pi0VsError}
\end{wrapfigure}
Figure \ref{fig:epoe_history} shows the history of results on the $\reepoe$.
The plot showing the final results is minuscule in the scale of early experiments.
Figure~\ref{fig:kl2pi0VsError} shows the error on the $\reepoe$ as a function of
the number $\klpiopio$ events ($N$).
The error is clearly proportional to \(1/\sqrt{N}\).
This means that the systematic errors were also reduced accordingly 
with statistical errors.
All these improvements in statistics and precision were made not only 
by the beam power.
Beam line design, chamber technology, photon measurement technology, 
trigger and data acquisition systems, and even magnetic tape technology
had to be improved along with it,
and behind them were people's innovative ideas, deep thinking, and many years of hard work.

%===============================
\section{Quest for $\kpinn$}
With the $\reepoe$ results and the observation of \CP\ violation in B decays,
the Kobayashi-Maskawa model was determined to be the source of 
\CP\ violations that had been observed in \emph{laboratories},
and became a solid piece in the \sm.
However,  the effect of this \CP\ violation mechanism is still too small to
explain the baryon -- antibaryon asymmetry in \emph{the universe}.
There should be new physics
beyond the \sm\ that violates \CP.
After the establishment of $\reepoe \neq 0$, 
kaon experiments changed their focus to search for 
new physics beyond the \sm.
To search for a small sign of new physics, there are two important points.
First, the background has to be small, and 
second, it has to be known with a small error.
Here, 
in addition to backgrounds caused by experimental techniques such as misidentifying particles or
mis-measurements, background decays caused by the \sm\ itself should also be well known.

%-------------------------------------------------------
\subsection{Physics of $\kpinn$}
\begin{wrapfigure}{r}{0.3\columnwidth}
        \includegraphics[width=0.3\columnwidth]{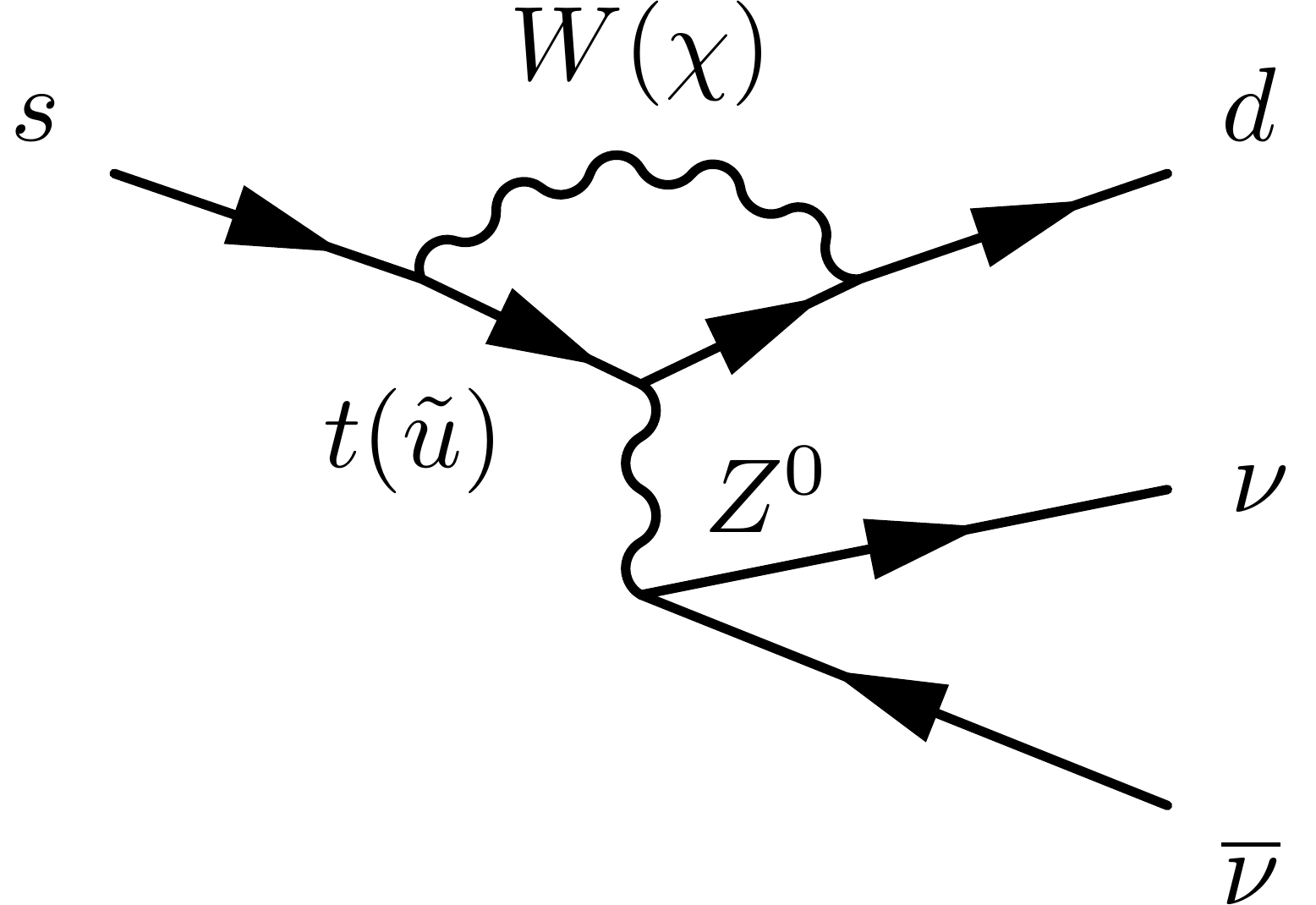}
        \caption{Penguin diagram of the $\kpinn$ decay.  In \sm, top quark is dominant in the loop.
        		New physics particles can enter the loop and have additional contribution.}
        \label{fig:pinn_penguin}
\end{wrapfigure}
The decay modes that have small and well-known branching ratios are
$\klpionn$ and $\kppipnn$.
These decay modes 
proceed through a penguin diagram as shown in Fig.~\ref{fig:pinn_penguin}.
In the \sm, the major contribution comes from a diagram with a top quark in the loop.
The branching ratios predicted by the \sm\ are small:
\(BR(\klpionn) = 2.4 \times 10^{-11}\), and
\(BR(\kppipnn) = 7.8 \times 10^{-11}\)~\cite{brod_prd83_2011}.
Also, theoretical uncertainties are only about $2-4$\%.
The current errors are dominated by the errors of the known CKM parameters, 
and those will be reduced in the upcoming B factory experiments.

New physics can contribute to these decays by having
new physics particles in the loop.
It can then change the branching ratios from the values predicted by the \sm.
The $\klpionn$ decay is sensitive to new physics that breaks \CP\  symmetry,
because $K_L$ is mostly \CP\  odd, and the $\pi^0\nu\overline{\nu}$ state is \CP\  even.

%-------------------------------------------------------
\subsection{History of $\kppipnn$ Experiments}
Let us first start from the charged $\kppipnn$ decay.
The signature of the decay is a single $\pi^+$ decaying from a $K^+$.
Major backgrounds are \(K^+ \to \mu^+ \nu\) decay where the $\mu^+$ is misidentified as $\pi^+$,
and \(K^+ \to \pi^+ \pi^0\) decay where the two photons from the $\pi^0$ are missed.

\begin{figure}[!ht] %  figure placement: here, top, bottom, or page
	   \centering
	   \subfigure[]{
		\includegraphics[height=0.32\linewidth]{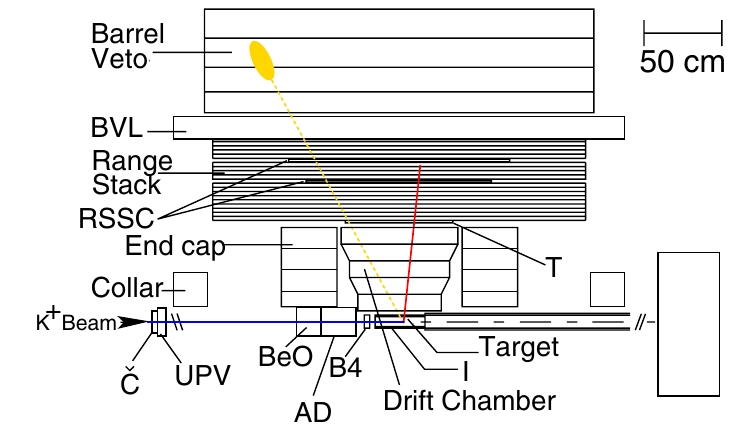}
    		}
    		\hspace{5mm}
	   \subfigure[]{
		\includegraphics[height=0.32\linewidth]{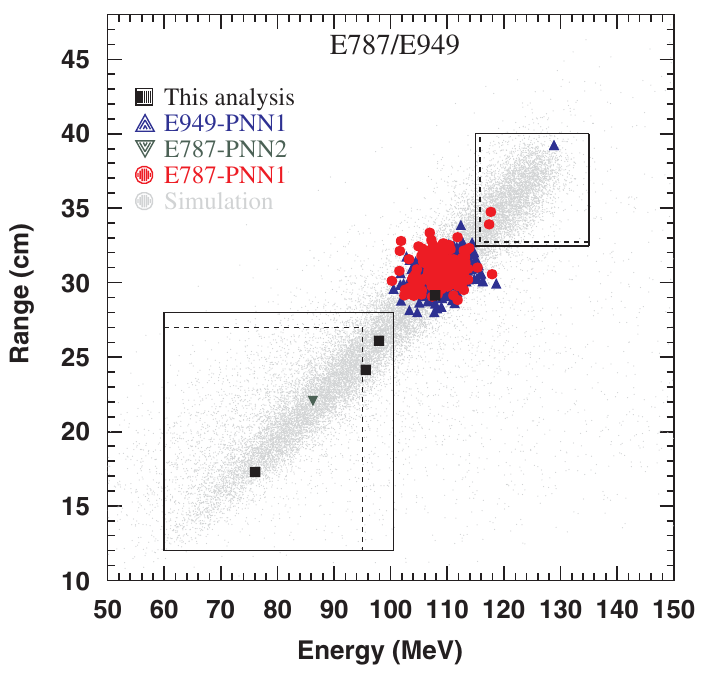}
    		}
	   \caption{(a) Schematic side view of the upper half of the BNL E949 detector.
	   (b) Range vs kinetic energy of all events passing all other cuts observed by 
	   the BNL E787 and E949 experiments~\cite{bnl949_prd79_2009}.
	   \copyrightAPS
	   }
	   \label{fig:bnl949}
\end{figure}

The search for $\kppipnn$ also has a long history, and dates back to 1970~\cite{klems_prl24_1970}.
The $\kppipnn$ events were first observed by BNL E787, 
and its branching ratio was measured by the E787 and E949 experiments.
Figure \ref{fig:bnl949}(a) shows the detector of BNL E949.
The experiment stopped the $K^+$ beam in a target, and looked for a single $\pi^+$ coming out 
from the target.
Stopping the $K^+$ simplifies the kinematics in reconstruction 
because the lab frame and the center of mass frame are the same.
The momentum of the $\pi^+$ is measured with a solenoid magnetic and a central drift chamber.
The energy deposit and the range of the $\pi^+$ were measured in a stack of scintillators (range counter) for identifying pions.
In addition, the \(\pi^+ \to \mu^+ \to e^+\) decay chain was traced by recording wave forms 
in the range counter to further identify pions.%
\footnote{This technique was first used in KEK E10 experiment to search for the $\kppipnn$ decay~\cite{asano_plb1107_1981}.}
The detector was surrounded by photon veto counters to suppress 
the background from \(K^+ \to \pi^+ \pi^0\) decays.
Figure \ref{fig:bnl949}(b) shows the scatter plot of the energy deposit and the range of 
$\pi^+$ events observed by BNL E787 and E949.
The experiments found 7 events in total, and gave 
\(BR(\kppipnn) = (1.73^{+1.15}_{-1.05}) \times 10^{-10}\)~\cite{bnl949_prd79_2009}.

Currently, CERN NA61 is preparing a new experiment to collect 45 $\kppipnn$ \sm\ events per year.
It uses high energy decay-in-flight $K^+$s to minimize hadron interactions in the beam line, 
and uses \v{C}erenkov counters to identify pions and kaons.
More details of the NA61 is covered by M.~Sozzi~\cite{sozzi}.

%-------------------------------------------------------
\subsection{History of $\klpionn$ Experiments}

Let us next move to the neutral $\klpionn$ decay.
Although the $\klpionn$ decay mode is theoretically clean, it is challenging experimentally.
One cannot trigger on the incoming $K_L$ because it is neutral, and the only observable particles 
are the two photons from the $\pi^0$ decay.
The initial $K_L$ momentum and its decay vertex is unknown, making the signal selection difficult.
In addition, there is a major background from the $\klpiopio$ decay mode if two of the four photons 
from the decay are missed.

The first experimental upper limit on the branching ratio was %of the $\klpionn$ decay was 
\(BR(\klpionn) < 7.6 \times 10^{-3}\) (90\% CL)~\cite{littenberg},
based on the Cronin's 
old result on $\klpiopio$~\cite{cronin_prl_18_1967}.
%The search for $\klpionn$ was first proposed by Littenberg~\cite{littenberg}.
%He even used Cronin's old result~\cite{cronin_prl_18_1967} to give the first upper limit,
%\(BR(\klpionn) < 7.6 \times 10^{-3}\) (90\% CL).

The first dedicated data was taken by KTeV E799-II, a rare-kaon experiment at Fermilab.
It had a one-day special run to look for events with only 2 photons which decayed from the $\pi^0$ in $\klpionn$ decay, and gave an upper limit, 
\(BR(\klpionn) < 1.6 \times 10^{-6}\) (90\% CL)~\cite{nakaya_plb_447_1999}.

The KTeV E799-II also
looked for the $\pi^0$ Dalitz decay.
Using the $e^+e^-$ tracks from the Dalitz decay, the decay vertex was reconstructed,  
the $m_{ee\gamma}$ was required to be consistent with the $\pi^0$ mass,
and the $\pi^0$ was required to have a transverse momentum $P_T > 0.16$ GeV/c.
Based on no observed events, the experiment gave 
\(BR(\klpionn) < 5.9\times 10^{-7}$ (90\% CL)~\cite{kazu_prd61_2000}.
Although the Dalitz decay can provide tight kinematical constraints, 
it has less sensitivity than the experiments with $\pi^0 \to \gamma \gamma$ decays,
because the \(BR(\pi^0 \to e^+ e^- \gamma)\) is only 1.2\%, 
and the acceptance for the $e^+e^-$ pair is small due to it small opening angle.

\begin{figure}[!ht]
    \begin{center}
        \includegraphics[width=0.77\columnwidth]{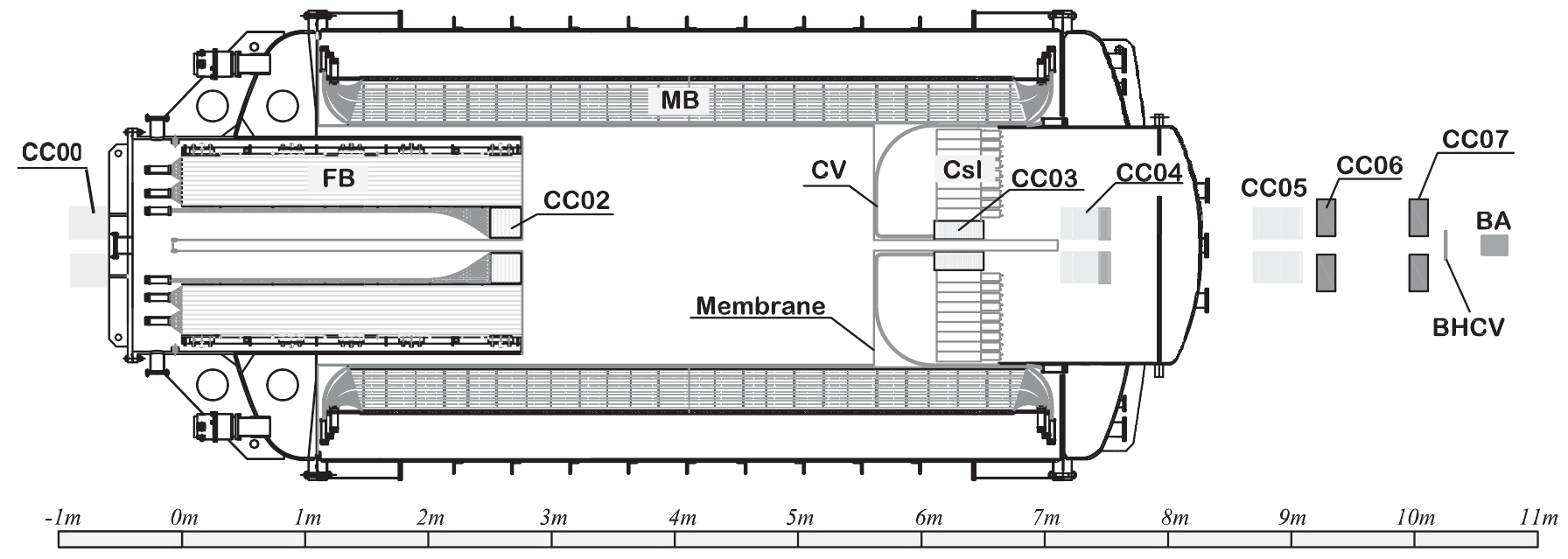}
        \caption{Schematic view of the KEK E391a detector
		(taken from \cite{e391a}).
		\copyrightAPS
		}
        \label{fig:e391a_detector}
    \end{center}
\end{figure}

The first dedicated experiment for the $\klpionn$ decay was KEK E391a experiment.
A $K_L$ beam with the average momentum of 2 GeV/c was made by bombarding a target
with 12 GeV protons.

\begin{wrapfigure}{r}{0.45\columnwidth}
        \includegraphics[width=0.45\columnwidth]{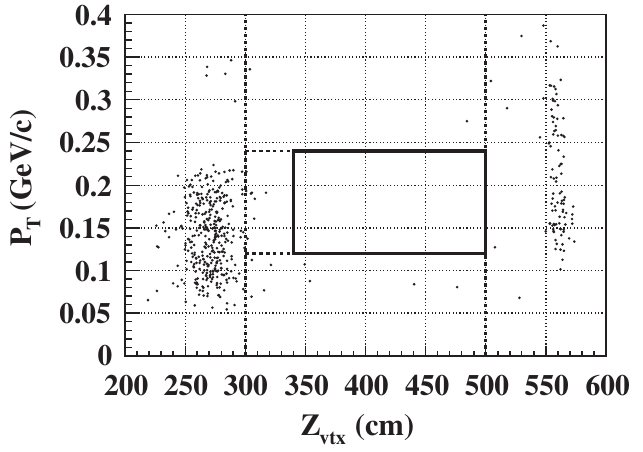}
        \caption{Scatter plot of the reconstructed $P_T$ vs the $Z$ position of the events that passed all other cuts
        		observed by KEK E391a experiment
		(taken from \cite{e391a}).  The solid rectangle indicates the signal region.}
        \label{fig:e391a_z_pt}
\end{wrapfigure}
As shown in Fig.~\ref{fig:e391a_detector}, 
the detector consists of an electromagnetic calorimeter to detect two photons from the $\pi^0$, 
and a hermetic photon veto system surrounding a decay volume to 
suppress the $\klpiopio$ background.
The calorimeter was made of 7-cm square and 30-cm long pure CsI crystals stacked inside a
cylinder with 1 m radius.
The surfaces of these detectors were covered by plastic scintillators to 
veto charged particles.
All of these detectors were housed inside a vacuum tank.
The neutral beam had to be in vacuum to suppress $\pi^0$s produced by 
neutrons interacting with residual gas, and thus required a beam pipe.
If the detectors were located outside the beam pipe, 
low energy photons could be absorbed in the beam pipe, and 
the $\klpiopio$ background would increase.
The solution was to minimize such dead material by placing most of the detectors inside 
effectively a large ``beam pipe''.
The calorimeter had a hole at the center for the beam to pass through.
In downstream, another photon veto counter was placed in the beam 
to veto photons escaping through the hole.

Figure \ref{fig:e391a_z_pt} shows the scatter plot of 
decay vertex and the transverse momentum ($P_T$) of $\pi^0$'s.
The decay vertex and $P_T$ were reconstructed by assuming that the two photons from a $\pi^0$
originated at the center of the beam area.
Based on no events in the signal region, the experiment lowered the upper limit to
\(BR(\klpionn) < 2.6 \times 10^{-8}\) (90\% CL)~\cite{e391a}.

%-------------------------------------------------------
\subsection{J-PARC KOTO Experiment}
A new $\klpionn$ experiment, called KOTO,  is starting up at the J-PARC laboratory in Japan.
It utilizes a high intensity 30-GeV proton beam to 
achieve a sensitivity close to the branching ratio predicted by the \sm~\cite{koto_proposal}.
The protons extracted from the Main Ring hit a common target shared by multiple experiments.
A neutral beam line is extracted at 16$^\circ$ from the proton beam line.
A new pair of collimators were designed to suppress neutrons in the beam halo.
The vacuum tank and photon veto detectors used at the KEK E391a were moved to J-PARC.
The electromagnetic calorimeter was replaced with the pure CsI crystals used for the Fermilab KTeV experiments.
With their small cross-sections (2.5-cm-square crystals in the 1.2 m $\times$ 1.2 m region, and 
5-cm-square crystals in the outer region), it has a better $\gamma$/neutron identification, and
it can suppress backgrounds caused by two photons hitting the calorimeter together,
which is another mechanism of missing a photon.
The energy resolution is also improved by their longer length, 50 cm (27$X_0$).
The charged veto counter covering the upstream side of the calorimeter was replaced with
two layers of plastic scintillators.  
Each plane is only 3 mm thick to suppress halo neutrons interacting in the counter.
The photon veto detector covering the upstream end of the decay volume was replaced with
CsI crystals to improve the veto efficiency and to detect beam-halo neutrons.
A new photon veto detector was placed in the beam downstream of the calorimeter 
to veto photons escaping through the beam hole in the calorimeter.
The detector was made of
modules consisting of a lead converter and an aerogel \v{C}erenkov counter
to have a low veto inefficiency ($10^{-3}$) for photons even under full beam intensity. 
The signals from all the detector components are digitized with flash ADCs.
They are used to record the waveforms to identify signals in a high-rate environment, 
and to produce trigger signals based on the digitized information.

The KOTO experiment started its first physics run in May 2013.
Although the run was terminated after 100 hours of data taking
due to a radiation accident in the experimental hall,
it is pursuing physics analysis intensively.\footnote{%
The first result was presented at the CKM2014 Conference at Vienna in September 2014.}
%Figure \ref{fig:koto_jan_3pi0_z} shows the decay z vertex distribution of 
%high statistics $\klpiopiopio$ events taken in January 2013;
%the Monte Carlo simulation reproduced the distribution well.
After the J-PARC Hadron Experimental Facility recovers from the accident, 
KOTO is planning to install a new photon veto detector, 
increase the beam intensity, and start high-sensitivity runs.

\subsection{Prospect of $\kpinn$ experiments}
\begin{figure}[!ht]
    \begin{center}
        \includegraphics[width=0.7\columnwidth]{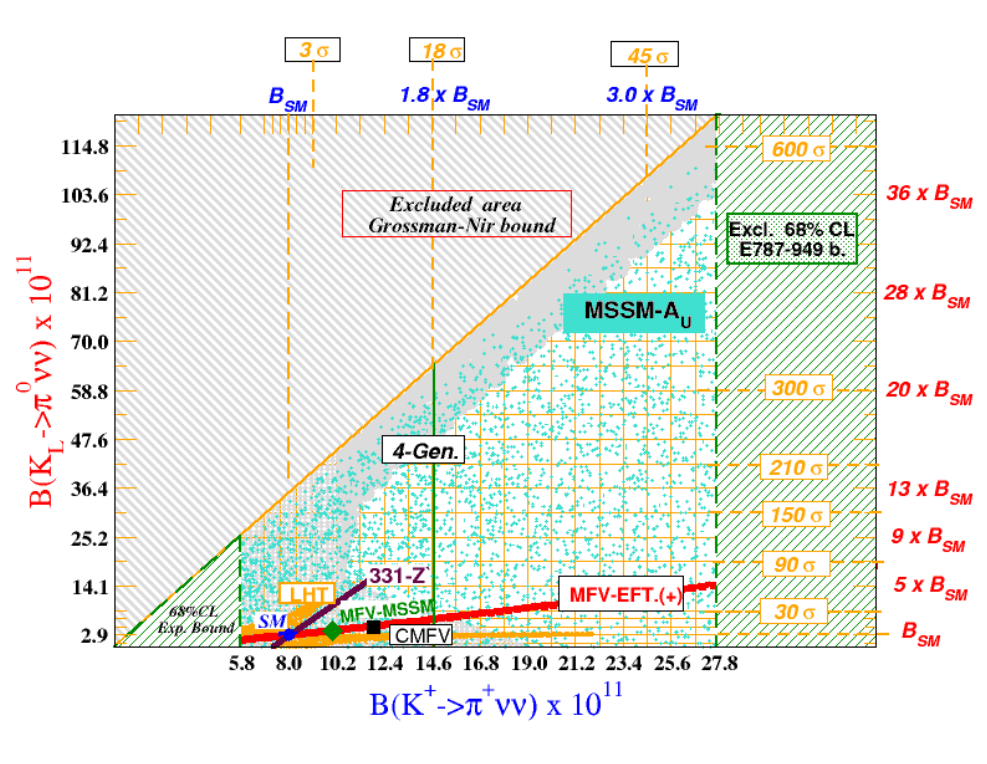}
        \caption{Branching ratios of $\klpionn$ vs $\kppipnn$ predicted by 
        		various theoretical models
		(taken from \cite{mescia}).}
        \label{fig:kpinn_br}
    \end{center}
\end{figure}

Figure \ref{fig:kpinn_br} shows the predictions of the branching ratios of 
$\kppipnn$ and $\klpionn$ by various theoretical models.
In the next few years, the CERN NA61 and J-PARC KOTO experiments will 
explore this large unexplored region.

%==============================
\section{Summary}
\begin{figure}[!ht]
    \begin{center}
        \includegraphics[width=0.7\columnwidth]{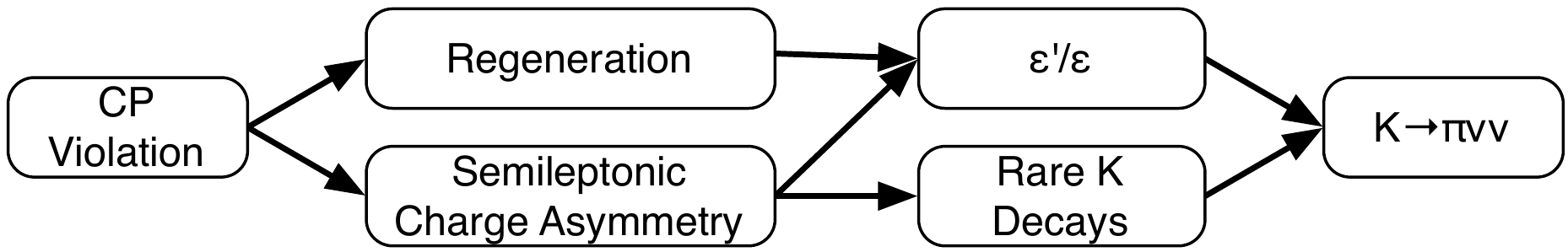}
        \caption{Evolution of kaon experiments since the discovery of \CP\  violation.}
        \label{fig:k_evolution}
    \end{center}
\end{figure}

Figure~\ref{fig:k_evolution} shows my view of how the kaon experiments have evolved in the past 50 years.
After the first series of \CP\  violation experiments in the 1960s, 
kaon experiments have once moved on to high precision experiments to measure the 
charge asymmetry in semileptonic decays and $K_S$ regeneration amplitudes.
The experimental technologies and knowledge accumulated then
later became the foundations of the modern $\reepoe$ experiments which established the non-zero value.
The stream of high-statistics experiments have also moved onto rare K decay experiments
which I could not cover in my talk.
These two streams are now recombined to the new $\kpinn$ experiments 
to search for physics beyond the \sm.
Now that we are finally entering an unexplored territory, we should be open to any new signs 
that may appear.

%\bigskip
\section{Acknowledgements}
I would like to thank the conference organizer for inviting me to give this review talk,
and thereby giving me a chance to study the great works of past which are 
the foundations of the current experiments.
This work was supported by JSPS KAKENHI Grant Number 23224007.

\end{document}